\documentclass[showpacs,showkeys]{revtex4}
\usepackage{graphicx}
\def\lsim{\mathrel{\rlap{\lower4pt\hbox{\hskip1pt$\sim$}}
    \raise1pt\hbox{$<$}}}         
\def\gsim{\mathrel{\rlap{\lower4pt\hbox{\hskip1pt$\sim$}}
    \raise1pt\hbox{$>$}}}         
\newcommand{\be}{\begin{equation}}
\newcommand{\ee}{\end{equation}}

\begin{document}

\title{Arguments for a ``U.S. Kamioka":  \\
SNOLab and its Implications for North American Underground Science Planning}

\author{W. C.  Haxton and K. A. Philpott}
\affiliation{Institute for Nuclear Theory and Dept. of Physics, University of Washington,
  Seattle, WA 98195}
\author{Robert Holtz}
\affiliation{Dept. of Civil and Environmental Engineering, University of Washington, Seattle, WA 98195}
\author{Philip Long}
\affiliation{Environmental Technology Directorate, Pacific Northwest National Laboratory, Richland, WA 99352}
\author{J. F. Wilkerson}
\affiliation{Center for Experimental Nuclear Physics and Astrophysics and Dept. of Physics, University of Washington, Seattle, WA 98195}

\begin{abstract}
We argue for a cost-effective, long-term North American underground
science strategy based on partnership with Canada, initial construction of a 
modest U.S. Stage I laboratory designed to complement SNOLab, and follow-up stages
to create clean horizontal access to greater depths.  
We show, by reviewing the requirements of detectors now in the R\&D phase,
that SNOLab and a properly designed U.S. Stage I facility would be capable of meeting
most needs of North America's next wave of underground experiments.

One opportunity for creating such a laboratory is the Pioneer tunnel in Washington State,
a site that could be developed to provide dedicated, clean, horizontal access.
This unused tunnel, part of the deepest (1040 m) tunnel system
in the U.S., would allow the U.S. to establish, at low risk and modest cost,
a laboratory at a depth (2.12 km.w.e., or kilometers of water equivalent)
quite similar to that of the Japanese laboratory Kamioka (2.04 km.w.e.).  
The site's infrastructure includes highway and rail access
to the portal, a gravity drainage system, redundant power, 
proximity to a major metropolitan area, and a system
of cross cuts connecting to the parallel Great Cascade tunnel and its
ventilation system.  We describe studies of cosmic ray attenuation important
to properly locating such a laboratory, and the tunnel improvements that
would be required to produce an optimal Stage I facility.

We describe the unique role this location could play in formulating an international
plan for high-energy accelerator physics that includes, as one component, a neutrino factory.
The site has a ``doubly magic"  baseline -- a 7500 km separation from
both KEK and CERN -- as well as an appropriate baseline for CP violation
studies, should FermiLab host the neutrino factory.

We also describe how new space at greater depth could be added in response to the needs
of future experiments, building on the experience gained in Stage I.
We discuss possible designs for Stage II (3.62 km.w.e.) and Stage III (5.00 km.w.e.)
developments at the Pioneer tunnel, should future North American needs for deep
space exceed those available at SNOLab.  This staging could be planned
to avoid duplication of SNOLab's capabilities while minimizing 
construction and operations costs.  We describe the existing
geotechnical record important to future stages, including
past tunneling histories,  borehole studies and analyses, and recent examinations 
of the Pioneer tunnel.  We also describe the significant broader impacts of this project in improving
the efficiency, safety, and security of one of the nation's key
transportation corridors.
\end{abstract}

\pacs{29.90.+r,14.60.Pq,26.65.+t,98.70.Sa}

\keywords{underground laboratories, cosmic ray muons, neutrinos, dark matter}

\date{\today}
\maketitle

\section{Introduction}
Some of the most compelling  questions in science -- the origin of dark matter, the nature of
neutrino mass, the stability of the nucleon, the source of the CP violation
responsible for the excess of matter over antimatter -- are motivating a new
generation of low-background experiments \cite{HS03}.  To escape interference from
cosmic ray muons and the secondaries they can produce, 
such experiments must be located deep underground.

Finding suitable space has been an important concern of underground scientists for
four decades.  European scientists have exploited the continent's many deep road and
railway tunnels. Italy's Gran Sasso \cite{GS}, a horizontal-access facility built off the 
Appenine road tunnel between L'Aquila and Teramo,  is perhaps
the premier European laboratory, providing 3.03 km.w.e. (kilometers of water
equivalent) of overburden.  [Throughout this
paper we define overbuden as the depth under a flat surface that would give an
equivalent muon flux, so that sites with different topographies can be fairly 
compared.  These depths are calculated in Ref. \cite{Philpott}.  Typically, for a mountain site, the peak overburden would be greater
than this depth by $\sim$ 0.6-1.2 km.w.e.]   Frejus, a laboratory built off the 
French-Italian road tunnel that connects Modane with Bardonecchia, is located
at a depth of 4.15 km.w.e.  Other European laboratories include Boulby (Great
Britain), CUPP (Finland),  and CanFranc (Spain).  Also notable is Russia's Baksan
Laboratory, for which a dedicated tunnel was constructed under Mt. Andyrchi in
the Caucasus, the first example of a purpose-built deep facility.

Japan has mounted a very successful underground program at Kamioka \cite{Kamioka},
a horizontal-access site within an unused portion of that mine.  Several large-volume
detectors for solar and atmospheric neutrino, nucleon decay, and reactor neutrino studies
have been deployed successfully there (2.04 km.w.e.).  Future
plans include a major Xe-based solar neutrino/dark matter experiment and a 
long-baseline accelerator neutrino experiment: T2K will
direct the neutrino beam produced by the Japanese Hadron Facility to the
50-kton Super-Kamiokande detector.

Lacking Europe's network of deep tunnels, North American scientists have 
fewer options for developing parasitic laboratories.  Three sites are 
currently in operation, and all require vertical (hoist and shaft) access, a feature
that frequently increases the cost and difficulty of underground operations.  Two of these
are in the U.S.  The Waste Isolation Pilot Project (WIPP) \cite{WIPP}, near Carlsbad, New 
Mexico, provides an overburden of 1.58 km.w.e., while Soudan \cite{Soudan}, a former iron
mine now operated by Minnesota as a state park, is at 1.95 km.w.e.  The former has a
modern high-capacity lift, but this lift is available to science only when
such use does not interfere with WIPP's main function.  
The latter has a hoist that is generally available for science, though the cage's 
internal compartment dimensions (1.2m by 1.8 m) and 
capacity (6 tons) somewhat limit access.  The Soudan Laboratory has conducted
a vigorous underground science program for many years, including dark
matter studies, long baseline neutrino physics, and proton decay. 

The third site is Canada's Sudbury Neutrino Observatory,
a laboratory built to house a heavy-water solar neutrino detector that had 
especially stringent background requirements.  SNO is located below 2 km
of rock in the Sudbury Mine, an active nickel mine.  
The site is now being developed as the world's first very deep (6.01 km.w.e)
multipurpose laboratory, dubbed SNOLab \cite{SNOLab}.  In our view SNOLab
is an important step forward that should influence the U.S.
strategy on underground science.

\subsection{U.S. underground laboratory planning and SNOLab}
In September 2000, recognizing the roles laboratories like Gran Sasso and Kamioka
had played in stimulating European and Japanese underground science,
U.S. scientists embarked on an effort to create a Deep Underground Science
and Engineering Laboratory (DUSEL) \cite{Seattle,Bahcall}.  This effort has a complicated history.  It
began before SNOLab was proposed and initially was focused on the
Homestake Mine, then an operating facility that appeared to offer an opportunity to quickly
establish a very deep laboratory  similar to SNOLab.   Despite strong support from
several community studies, DUSEL ran into a number of obstacles.  The end result
was a March 2004 offer by the National Science Foundation, the agency charged with
considering DUSEL, to restart the process.  Eight interested site groups were invited
to submit preproposals requesting funds to develop conceptual proposals.
In July 2005 the NSF decided to provide conceptual design funding to two sites,
Henderson and Homestake, both of which are
mines with vertical access, like SNOLab.   The Foundation has recently announced that
the next step of competition, called S3, will be open to all interested sites, which will
allow purpose-built drive-in access designs to be evaluated, as well.

In our view SNOLab is important to the strategy the U.S. follows in underground science.  SNO and
SNOLab make use of one of the world's deepest continuous straight shafts, the 7138-ft
Creighton \#9 shaft.  This allows experimenters to transport scientific packages from the
surface directly to the 6800-ft level, where they are moved by rail 1.5 km to the
laboratory location.   The mine environment and hoist constraints 
have led to a ``blue box" mode of experiment construction:
loads of up to  3.5m by 1.1m by 1.7m can be moved underground in these boxes.
The science use is economical
because the infrastructure cost of vertical access is borne primarily by the mine owner.  The SNOLab
expansion will provide space for several new experiments, a sophisticated surface
laboratory with support facilities, and improved facilities for maintaining the clean
barrier between experiments and mining activities.
SNOLab will be completed in 2007.

There are important similarities and some differences between SNOLab and the U.S. mine
sites, Henderson \cite{Henderson} and Homestake \cite{Homestake}.  Henderson's 
hoist and 8.5m-diameter concrete-lined shaft are high capacity, capable of handling
loads larger than those possible at SNOLab.   But
the shaft reaches only 945m underground.  Thus, to achieve DUSEL depths, very
significant new construction is required:
existing mine drifts would be extended by the excavation of approximately 7 km 
of access tunnels and 7 km of ventilation drifts, to reach an elevation of 4900 ft (the deep
campus) under Harrison Mt., where the overburden would be 5.29 km.w.e.  The character of
Henderson operations will also change during DUSEL's expected lifetime:
while initially DUSEL-Henderson will be a parasitic facility 
like SNOLab, mining is expected to cease while
DUSEL is still operational.  Henderson's molybdenum ore body will be exhausted in about 20-25
years, 10 to 15 years after DUSEL opens.  When mining operations terminate and science
becomes the sole tenant, considerable effort will be needed
to reduce the mine's footprint and mechanical systems in order to make 
subsequent operations economically feasible.  Henderson ranks among the world's
ten largest mines.

Homestake would provide great depth, like SNOLab, but with important differences
in the access.  First, Homestake requires two steps to reach depth, use of the 
Ross or Yates shafts/hoists to the mine's 4850 ft level (4.16 km.w.e.), and then either
the \#6 or \#4 shafts and winzes to the proposed DUSEL
site at the 7400-ft level (6.43 km.w.e.).  All four hoists are
necessary because dual access underground must be maintained.
Second, because commercial operations ended in 2002, DUSEL would be the sole
tenant of the mine.    In a dedicated vertical-access science laboratory of this
nature, non-science facility operations costs would be quite high, dominating facility lifetime costs,
according to existing studies.  (See, for example, \cite{HS03} and the discussion of Sec. IIIc.)   
However, while a dedicated vertical access laboratory is more
costly than a parasitic vertical laboratory, it provides hoists that can be used
exclusively for science.

In 2003 a  community collaboration published an engineering plan \cite{HS03} for converting the 
Homestake Mine to DUSEL, meeting (to the extent possible) the technical
design criteria established in the Bahcall Report \cite{Bahcall} and including surface
facilities comparable to those of Gran Sasso.  The construction and associated 5-year 
operations costs for DUSEL-Homestake 
were estimated to be  \$301M (FY03, excluding the education/outreach construction/operations
components of the proposal).  The plan included the 
rehabilitation and modernizing of the shafts and hoists, to lower operations costs and
provide a usable hoist platform of 3m by 3.5m -- part of the difficult task of converting
a large industrial facility into a smaller, stand-alone science laboratory.
It assumed an operating mine would be turned over to science.  Today this plan would have to be  
modified because most of the underground systems have been dismantled,
surface facilities such as the hydroelectric and water processing plants are no longer available, and 
the mine's flooding has progressed to about the 6000-ft level.

\subsection{The Pioneer tunnel, SNOLab, the neutrino factory, and international cooperation}
A number of conditions have changed since DUSEL was first discussed in September 2000.
At that point SNOLab had not been proposed, so that no new deep underground space was
on the North American horizon.  The U.S. was enjoying a 
budget surplus.  During the DUSEL process both NSF and community studies 
noted the attractiveness of horizontal sites, but
no existing openings of this type had been identified. 

In August 2005 the owner of the Pioneer tunnel, after reviewing an engineering report \cite{SW2}
addressing scientific use of the tunnel,
concluded that such use would not adversely impact railroad operations in
the neighboring Great Cascade tunnel.  The Pioneer and Great Cascade tunnels 
form the  longest (12.5 km) and deepest (1040 m)  system in the U.S.
The Pioneer tunnel could be adapted
to provide the U.S. with a drive-in facility very similar in depth to Japan's national laboratory
at Kamioka, but with dedicated and exceptionally clean entry.  The site's attributes
include:\\
$\bullet$ The Pioneer tunnel is available now, and for the foreseeable future.
As no new tunnel construction is required, technical risks in
establishing a laboratory are minimal. \\
$\bullet$ The tunnel is 1.5 hours (75 miles) from a major metropolitan area with a high
concentration of high-tech industry (Seattle), and from Puget
Sound's international airport and shipping ports.  The portal is directly accessible by both 
highway and railway.\\
$\bullet$ The horizontal access at modest positive gradient (1.56\%) would allow large loads of the type described in the Bahcall Report (20-ft cargo containers) to be transported underground by truck or rail.\\
$\bullet$ The site is privately owned by Burlington Northern \& Santa Fe (BNSF).  As the site is currently
permitted for railway-related activities similar to those proposed for science (e.g.,
drainage, ventilation, and a variety industrial activities in the portal
area), the additional permitting required for the laboratory should be straightforward.\\
$\bullet$ The site provides the safety of a dual-bore system.  Approximately 30 crosscuts
connect the Pioneer tunnel to the Great Cascade tunnel/ventilation system.\\
$\bullet$ Redundant stable power is available at the portal, provided by independent transmission lines
that connect directly to 
two Columbia River dams.\\
$\bullet$ The ambient rock temperature at the experimental site is 21$^\mathrm{o}$C.\\
$\bullet$  The site is unique in providing ``magic" baselines to both CERN and KEK and a nearly 
optimal CP-violation baseline for a FermiLab neutrino factory.\\
$\bullet$ A potentially interesting earth science program could be conducted in 5 kilometers 
of ventilation tunnel that we propose to leave unlined.\\
$\bullet$ In the future, the site could be further developed in successive stages, should a need arise, 
to provide clean drive-in access to locations with significantly greater overburdens.

We believe the Pioneer tunnel could play an important role in a staged,
long-term strategy to make North American underground science competitive
internationally.   Our views have been influenced by a strategy our European
colleagues are exploring.  Building on the success of Gran Sasso, government
science agencies of France and Italy, DSM (Direction des Sciences de la Matiere),
IN2P3 (Institut National de Physique Nucleaire et de Physique des Particules),
and the INFN (Instituto Nazionale di Fisica Nucleare), have entered into an 
agreement to create a joint Frejus-Gran Sasso European Underground Facility.
The agreement envisions a major expansion of Frejus designed
to complement, rather than compete with, Gran Sasso.  In this way the two nations
and two laboratories could work together to meet European underground
science needs.

In our view, there are even greater opportunities for successful international
cooperation between Canada and the U.S., due to the natural
complementarity of two sites we could develop:\\
$\bullet$  Establish at the existing moderate depth of the Pioneer tunnel a Stage I facility
-- a U.S. Kamioka -- providing dedicated horizontal access, portal-to-depth cleanliness, and
outstanding excavation capability.   This facility could
be developed quickly.\\
$\bullet$  Establish a cooperative agreement with SNOLab that allows the
U.S. and Canadian funding agencies to work together in supporting and siting experiments.  Those
experiments that require great depth and are compatible with SNOLab's ``blue box"  access should be
sited at SNOLab.  Experiments requiring only intermediate depth or with needs not
easily satisfied at SNOLab (rail or truck access for economic construction,
separation from other experiments or commercial operations, long baselines,
location within the US because of security requirements) could
be sited at the Pioneer tunnel (or at WIPP or Soudan).  We argue in this paper
that Stage I and SNOLab can meet the needs of next-generation North American
experiments.\\
$\bullet$  Because Stage I would be inexpensive to construct and very
inexpensive to operate, the U.S. would be able to direct the majority of currently available
funding to new experiments, helping to ensure that Stage I, SNOLab, and existing
U.S. underground facilities are fully utilized over the next decade.\\
$\bullet$  With the experience gained in constructing and operating Stage I,
the U.S. could turn to Stage II development of the Pioneer tunnel when
the need for additional deep space becomes apparent.  We believe such a
pragmatic, step-by-step approach would help the U.S. produce an optimal
facility.  A cooperative program based on SNOLab and Stage I/Stage II
would be very competitive with the joint facility France and Italy have proposed, and
would increase the range and number of experiments that North American scientists
could mount.

This site could also play a special role in international planning for high energy 
accelerator facilities, because of neutrino factory proposals.  The scale of
facilities in high-energy physics is now such that new projects must be undertaken by
international collaborations.  Specifically, the recently released National Research Council blueprint
for the US community, EPP2010 \cite{EPP}, designated the International Linear Collider (ILC) as
the next major experimental facility that should be built, and urged the US to
bid to be the host country.  This report also advocated moving neutrino physics
forward through a well-coordinated staged program of experiments developed with international
planning and cooperation.

A neutrino factory producing intense beams may be essential to future efforts to determine neutrino 
mixing angles, measure leptonic CP violation, and determine the mass hierarchy.
The recent APS Multidivisional
Neutrino Study \cite{APS} noted that the optimal baselines for neutrino-factory experiments are rather
well defined:  measurements done at the ``magic" baseline of 7500 km will allow experimeters
to determine $\theta_{13}$ and the mass hierarchy, free of uncertainties from CP violation.
The study pointed out the importance
of a followup experiment with a baseline of 3000 km to measure CP violation.

It would be helpful to neutrino factory aspirations to develop underground sites that would
meet these baseline requirements.  As Huber and Winter have pointed out \cite{huber}, 
there are only two locations that are guaranteed to have appropriate baselines,
for each of the three plausible neutrino factory sites (FermiLab, CERN, and KEK).  One
of these is in far eastern Russia, where no underground facilities exist; the other is very
near the Pioneer tunnel.  The Pioneer tunnel is ``doubly magic," with baselines to KEK
and CERN of 7240 and 7780 km, respectively.  This fortunate
circumstance could facilitate agreements where
our international partners would participate in the construction and science of a
U.S.-based ILC, while the U.S. reciprocates by supporting a KEK or CERN neutrino
factory through construction of a magic-baseline detector at the Pioneer tunnel.
Conversely, if a decision is reached to build the ILC at an off-shore location
like CERN, this relationship would then be reversed.
FermiLab's future would depend on neutrinos and/or muons.
If an underground facility existed at the Pioneer Tunnel, FermiLab would be remarkably
well positioned to move forward on a neutrino factory:  Gran Sasso, Europe's leading
underground laboratory, is magic (7330 km), while the Pioneer tunnel provides the needed
far-west-coast site optimizing the CP-violation experiment (2630-km baseline). 

This paper is arranged as follows.  In Section II we describe the history that led to
the identification and established the availability of the site.  We describe the physical
setting and the studies of cosmic ray attenuation we have performed, taking into detailed account
the mountain topography.  The resulting contour maps of cosmic-ray muon fluxes determine 
the optimal locations for Stage I and future
laboratory development.  We also describe the site's extensive geotechnical database
due to past construction, past borehole studies, and recent tunnel inspections.
In Section III we present a possible plan and estimate the cost of the
access improvements we would recommend.  This plan involves fully finished 
entrance-tunnel and laboratory space (so that surface-laboratory-quality conditions are maintained
from portal to depth), a separate ventilation/utility tunnel, and outstanding safety because
of crosscuts to the parallel Great Cascade tunnel and ventilation system.  We describe why this plan is potentially
important to other parties, including BNSF, Amtrak, and Puget Sound port districts,
due to improved safety, 
drainage, and throughput on one of the nation's most important rail transportation
corridors.  We also point out the critical role this site might play in helping address container
security issues.  In Section IV we review the requirements of key experiments that
the North American science community hopes
to undertake in the next 10 to 15 years, assessing how well the needs of these
experiments could be met by a combination of SNOLab and a ``U.S. Kamioka."
We discuss some of the special circumstances (rail support, rock quality, location)
that make the Pioneer tunnel site
especially attractive as a location for megadetector construction. 
In Section V we describe Stage II (and Stage III) possibilities for expanding the laboratory,
when a need develops for deep space beyond that available in SNOLab.  
Stage II, which would build on the access, electrical, drainage, and haulage systems
implemented for Stage I, would
provide, at modest cost, a drive-in laboratory with a cosmic ray flux nearly three times
lower than that of Gran Sasso.  In
Section VI we present some concluding remarks.

\section{The Pioneer Tunnel and Mt. Stuart Batholith}
We summarize the history of the Pioneer tunnel site.  In late 2003 a systematic national search was done
to identify sites that could be developed into a horizontal-access laboratory
like Gran Sasso, but with significantly greater depth. The advantages of the ``Gran Sasso model''
were summarized in the Bahcall Report and in a 2003 NSF Site Panel study:\\
$\bullet$ Horizontal sites provide easier access, simplifying construction and lowering the
costs of experiments.\\
$\bullet$ Horizontal sites require less investment in permanent and temporary infrastructure (hoists, pumps, etc.). \\
$\bullet$ Such sites have significantly lower operations, manpower, and maintenance requirements,
factors that otherwise can dominate lifetime facility costs.

Among the site candidates identified, perhaps the most remarkable was the Mount Stuart batholith in
the Cascade Mountains, Washington State.   The attractive properties of this
granodiorite and tonalite mass, 600 km$^2$ in
extent, include the known rock quality (ideal for excavation by tunnel boring machine),
low uranium content, and high relief primarily glacial in origin.  But the
site's most important attribute is its
tunneling history.  The batholith is home to four minor 
and three major tunnels, including two that comprise the longest and deepest tunnel
system in the U.S., the Cascade and Pioneer tunnels.  The 21 km of tunneling for this system
was completed in less than three years (1926-29).  The 11m/day average advance
rate achieved in the Cascade tunnel (before modern drilling/blasting techniques and
long before tunnel boring machines) was among the hard-rock tunneling records set
during the construction.

When the National Science Foundation announced, in March 2004, its intent to restart the
DUSEL process, a group formed to request NSF funds to support a conceptual design 
proposal for the batholith, selecting a site near Cashmere Mt. \cite{DC}.  
A geotechnical evaluation \cite{SW1} of the
batholith was conducted by the engineering firm of Shannon \& Wilson, Inc., in support
of the funding request.  Because the Pioneer tunnel is unlined and penetrates the
same rock mass, a major portion of the
geotechnical report focused on an examination of the Pioneer tunnel to assess
the long-term behavior of lightly supported openings 1000m
below the batholith's surface.

The report pointed out that the Pioneer tunnel was
in good shape, unused, and thus itself potentially interesting for science.  In October 2004
a representative of the University of Washington 
approached the tunnel's owner, Burlington
Northern \& Santa Fe, about tunnel use.  The company was receptive, but asked the
university to conduct an engineering study of implications of such use for the
railroad, utilizing firms familiar with tunnel operations (Shannon \& Wilson, Parsons Brinckerhoff
Quade \& Douglas).  The possible availability of this site was first
made public in February 2005, when it was included as the second site in University of Washington
preproposal requesting funds for conceptual design of DUSEL-Cascades.

The requested engineering study \cite{SW2} was completed in April 2005 and submitted to BNSF.
In August 2005 BNSF responded favorably, indicating that dedicated science use of the
Pioneer tunnel was compatible with BNSF operations in the main (Cascade) tunnel,
provided the new use is sequestered from BNSF operations and confined to areas south
of the Pioneer Tunnel.  BNSF also identified reasons such use could be of benefit to
the railroad, as discussed in Section III.   This paper reports the results of studies
of the tunnel alignment, cosmic ray muon shielding, and possible laboratory designs 
that have been completed since site availability was established.

\subsection{Tunnel alignment and geology}
The tunnel alignment is shown in Fig. 1.  The western portals for the Great Cascade and Pioneer tunnels
are located in the small town of Scenic, 75 miles from Seattle, at an elevation of 685m.    
As Scenic is west of Stevens Pass, on the Seattle side
of the Cascades, the route from there to SeaTac International Airport is entirely at
low elevations, on major highways (Interstate 5 and Highway 2).
The Great Cascade tunnel runs eastward for 12.5 km, 
beneath the Stevens Pass Ski Area, surfacing again
at Berne.  The Pioneer tunnel follows a parallel path, running along the south side of
the Great Cascade tunnel for 8.6 km, separated from the main tunnel by 20m (measured
centerline to centerline).  Its eastern terminus is underground, at 
a point of minimum overburden under Mill Valley, where a  200m  shaft
connects the tunnel to the surface.  The shaft is now closed and 
and capped, but could be reopened in the plan discussed in Section III,
to serve as an intake for ventilation and power.

\begin{figure*}
\includegraphics[width=18cm]{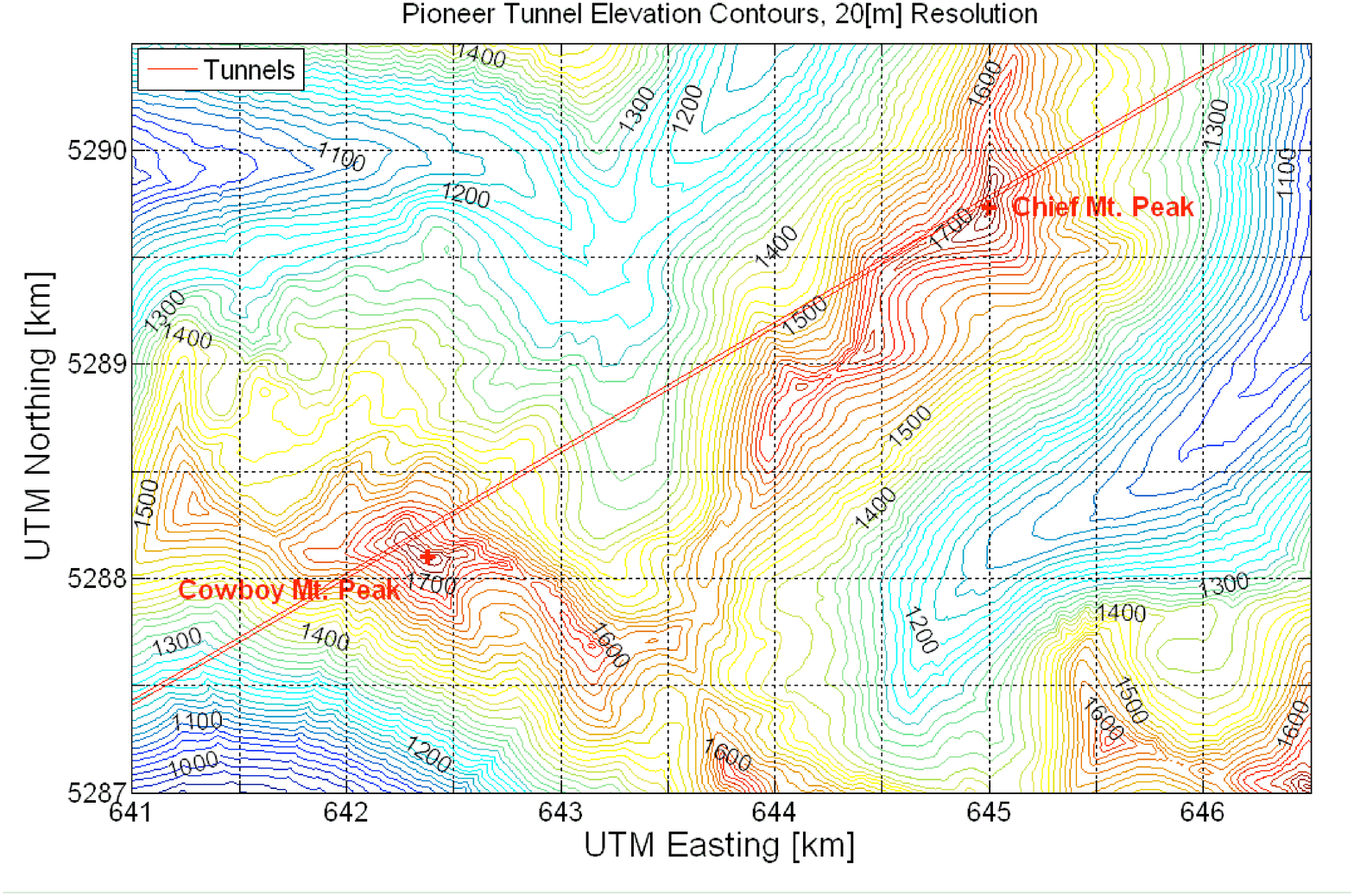}
\caption{The alignments of the parallel Great Cascade (north) and Pioneer tunnels
are superposed on a topological map of the Stevens Pass area of the Mt. Stuart batholith.}
\end{figure*}

The tunnel passes almost directly below the two tallest peaks in the Stevens Pass area, 
Cowboy Mt. and Big Chief Mt., both approximately 1785m in elevation.  This places the tunnel
very close to the points of greatest overburden.  The proposed Stage I laboratory location is under
Cowboy Mt., in an area immediately south of the Pioneer Tunnel,
3.8km from the western portal at Scenic and 4.8 km from the Mill
Valley shaft.

Figures  2 and 3 show a section
of the surface geologic map, developed by Tabor et al. \cite{Tabor}, and a cross section 
of the batholith along the tunnel alignment, with the proposed locations of Stages I and II shown.
Cowboy Mt. is near the northern end of the Mt. Stuart batholith, which intruded
Chiwaukum Formation schist and associated banded gneiss about 90 million years ago.
The rock type in the Cowboy Mt. area is granodiorite. 
The geology, hydrology, and support has been mapped in detail
along the tunnel: the resulting charts are available in Ref. \cite{SW1}.  A comparison
of the geologic maps of surface exposures with tunnel geology indicate rock conditions
within the batholith are relatively consistent over distance and depth.
 
As discussed in Section III, the
western 3.7 km of the tunnel will serve as the personnel and equipment entrance to the 
laboratory area beneath Cowboy Mt.   The tunnel, while primarily in the granodiorite,
also penetrates a zone of older biotite schist that
has undergone significant metamorphosis.  The zone is about a kilometer in width.
The proposed Stage I laboratory location is about a kilometer beyond the eastern
edge of the schist, embedded deep in the granodiorite, the preferred rock type for
laboratory construction.

\begin{figure*}
\centering
\includegraphics[width=17.5cm]{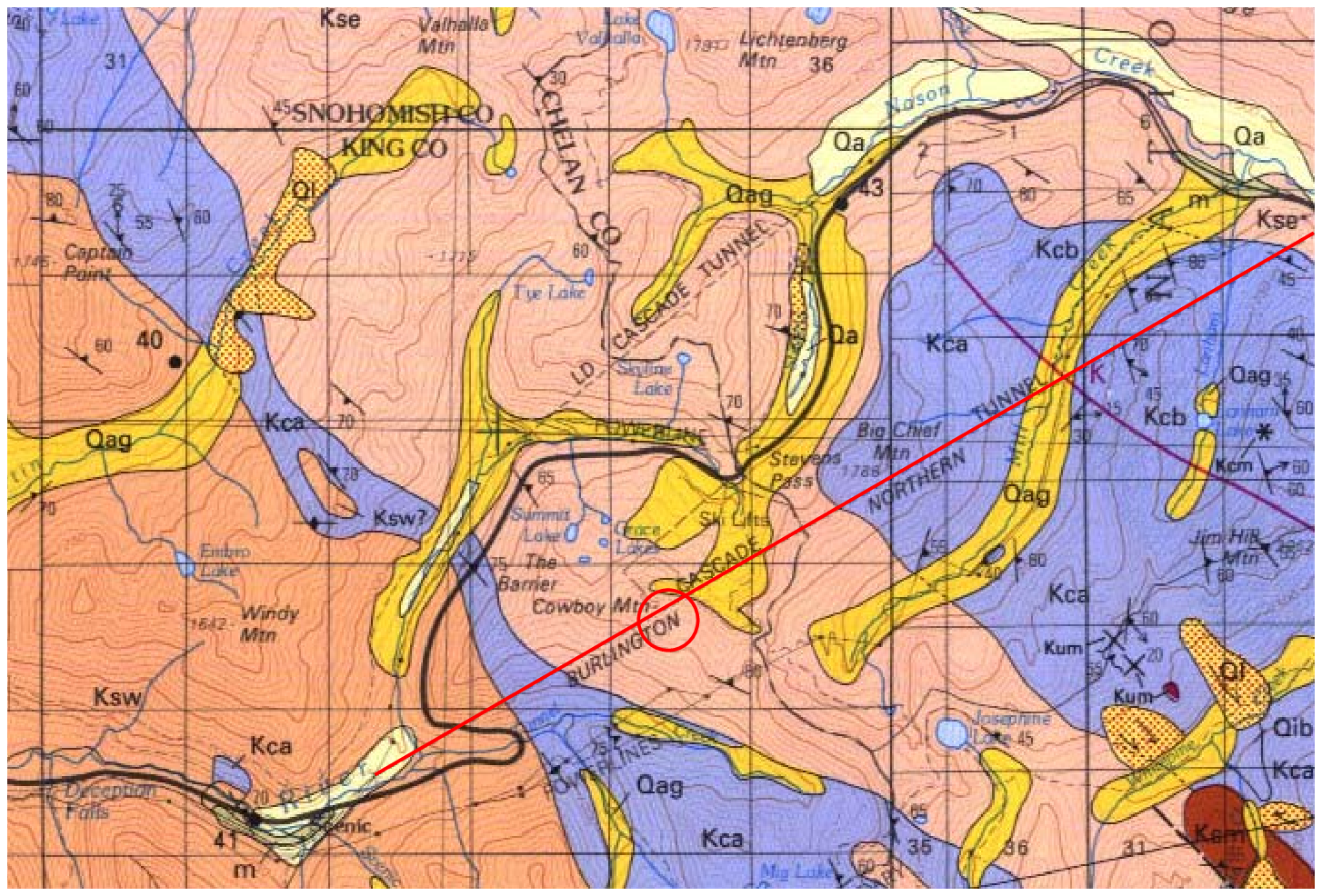}
\caption{The geologic map \cite{Tabor} of the region of the Mt. Stuart batholith penetrated
by the Great Cascade and Pioneer tunnels.  The tunnel alignment and the encircled
area around Cowboy Mt., the proposed laboratory location, are both highlighted in red.}
\end{figure*}

\begin{figure*}
\includegraphics[width=18cm]{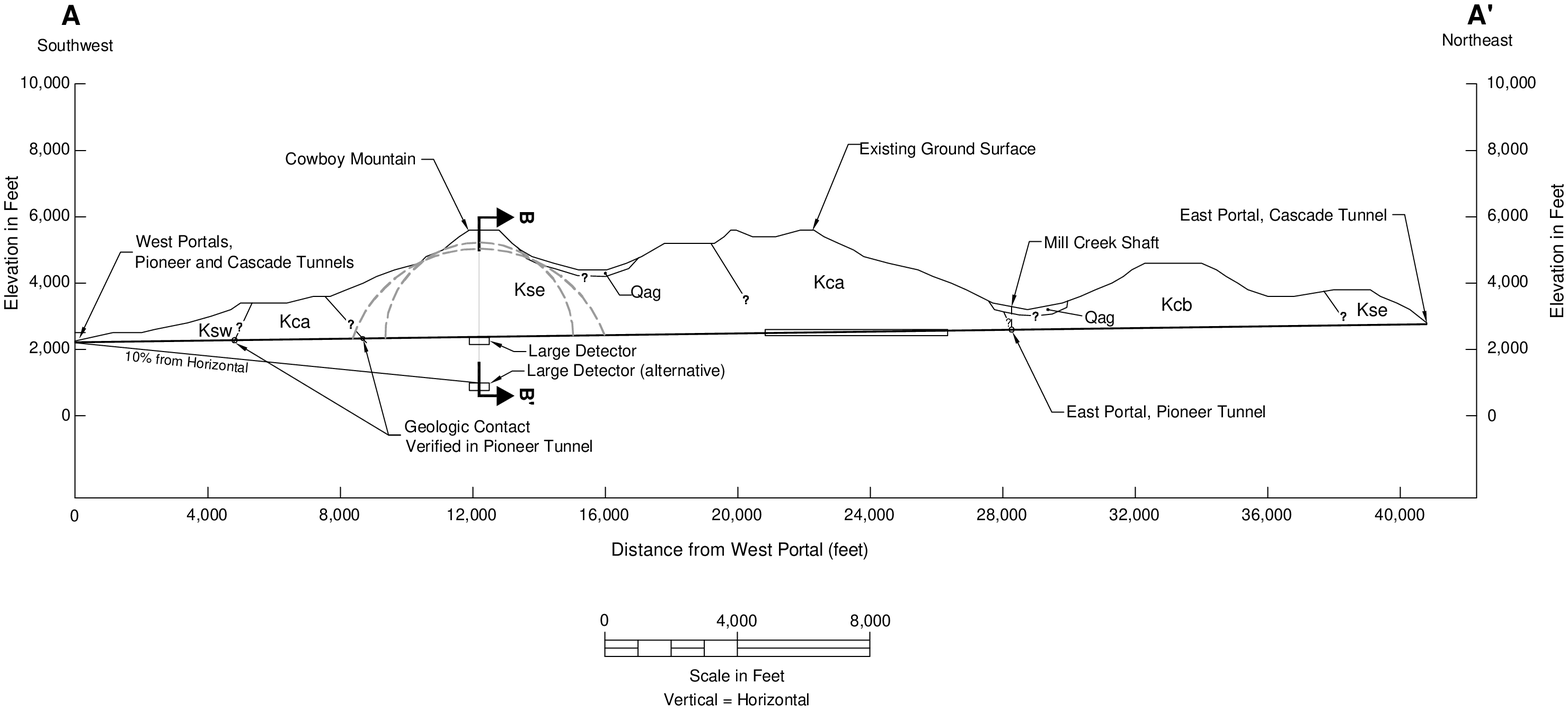}
\caption{A cross section of the batholith along the Pioneer tunnel.  The labels
``large detector" show the positions of Stage I and II.}
\end{figure*}

The 9 ft-wide, 8-ft high Pioneer tunnel was excavated by drill-and-blast methods.
Its purpose was to provide access for multiple construction headings on the main 
(Great Cascade) tunnel, and to provide a long-term drainage gallery.  In addition, 30 crosscuts,
separated on average by about 300m, and 14 refuge bays were constructed between the two
tunnels, so that the Pioneer tunnel could serve as an escape route in case of problems in
the main tunnel.  

The last detailed inspection of the tunnel was carried out in 1998 by geologists from
Shannon \& Wilson.  A return visit was made in 2004.  The exposed tunnel walls generally
consist of good to excellent rock, though with shear zones at points where the granodiorite
intersects layers of metamorphic rock.  These are regions where groundwater seepage is
prominent.  Total drainage from both tunnels (21 km  in total) was found to be moderate, 600 gpm
(gallons per minute).  
In the granodiorite, approximately 80\% of the tunnel
is unlined and unsupported.   Concrete sets or cast-in-place concrete was used
in 13\% of the granodorite sections and timber sets in 7\%.  In the schist, 
approximately 65\% of the tunnel
was left unlined, 28\% is concrete lined, and 8\% is timber lined.  This indicates that the
schist requires somewhat more support, and that granodiorite is the preferred rock
for laboratory excavation.  The general conclusion that the rock is of good quality
is consistent with the construction records for the tunnels, which note that the rock was
conducive to excavation by drill and blast, 
allowing rapid advance of the main tunnel.  It is
also consistent with the generally good condition of the tunnel after 80 years.

Joints  were evaluated as slightly rough, tight, and slightly altered.  Joint spacing in the
granodiorite ranged from 5 ft to 10 ft and higher.  In the schist the spacing ranged from 2 ft
to 5 ft or more.  Samples of granodiorite from the tunnel walls were tested
in the laboratory, yielding uniaxial compressive strengths (UCS) of 15 kpsi, 
a Rock Quality Designation (RQD) of 85\%, 
a Tunneling Quality Index Q of 21.3, and a Rock Mass Rating (RMR) of 62.  These ratings
are in the good category.  As discussed in Ref. \cite{SW1}, one expects a UCS of
approximately 25 kpsi to be typical of Mt. Stuart batholith granodiorite.
A schist sample was tested at 20 kpsi UCS.

Rock density and U/Th measurements were made for  Mt. Stuart batholith rock samples taken from
the slopes of Cashmere Mt., the primary region of study in the DUSEL-Cascades effort.
Densities ranged from 2.69 to 3.04 g/cm$^3$.  Many of these samples
had undergone significant weathering.  We use a density of 2.9 g/cm$^3$ in the
calculations reported below.  U/Th measurements were made for four samples \cite{Miley}.  Average
U/Th values for samples that may be typical of Pioneer tunnel depths are 0.77/0.53 ppm.  
Origin of the batholith from chemical fractionation of large (hidden) volumes of
relatively mafic parent magma is apparently responsible for its generally low U/Th
concentrations \cite{erikson}.

The geotechnical database includes several papers based on coring
studies that were performed by geologists associated with the Forest Service and USGS.   
One 102m coring 
was taken about 550 m south of the portal at Scenic and a second 152m coring
at a point 1.7 km to the east,
approximately 420 m south of the Pioneer tunnel.  This second coring probes an area
that would be about half way along the likely alignment of a new tunnel, should Stage II
be undertaken some day.  The geologic discussions in the papers are basic and consistent
with surface and Pioneer tunnel geology.  The primary motivation for the corings was to probe the
thermal gradient in this region of the batholith: there are geothermal springs near
Scenic.  The results show that the gradient is elevated locally in the area around Scenic,
but relaxes rather quickly as one moves away.   The results are consistent with the
thermal gradient in the Cowboy Mt. area deduced from Cascade tunnel construction records of about
16$^\mathrm{o}$C/km.  The rock temperature of new openings at the proposed 
Stage I laboratory location is 21$^o$C.

\subsection{Cosmic ray muon studies}
The August 2005 expression of support by BNSF for science in the Pioneer tunnel led
us to extend Pioneer tunnel studies to cosmic-ray muon attenuation.   In mountain topographies
meaningful depth determinations require calculations of muon flux that
account for the irregular surface: naive estimates based on peak overburdens are 
unreliable and generally overly optimistic.  Such calculations are also
important in properly positioning laboratory facilities.

The topography can be taken from standard 10m Digital Elevation Models (DEMs) available
from the USGS and other sources.   The muon flux at depth depends on the surface muon spectrum,
which is determined by the primary cosmic ray flux, the amount of atmosphere penetrated,
and small corrections due to muon energy loss in the atmosphere.  The increase in the
effective atmospheric slant depth with zenith angle $\theta$ is important, leading to a
flux that is sometimes parameterized as growing with $\sec \theta$.  Once the surface
flux is determined, the flux at depth is determined by the rock slant depth (a function of
$\theta$ and azimuthal angle $\phi$, through the topography), which governs muon energy
loss.  Effectively, the energy loss over the rock slant depth determines the minimum muon energy
for survival to the laboratory location.

\begin{figure}
\includegraphics[width=9cm]{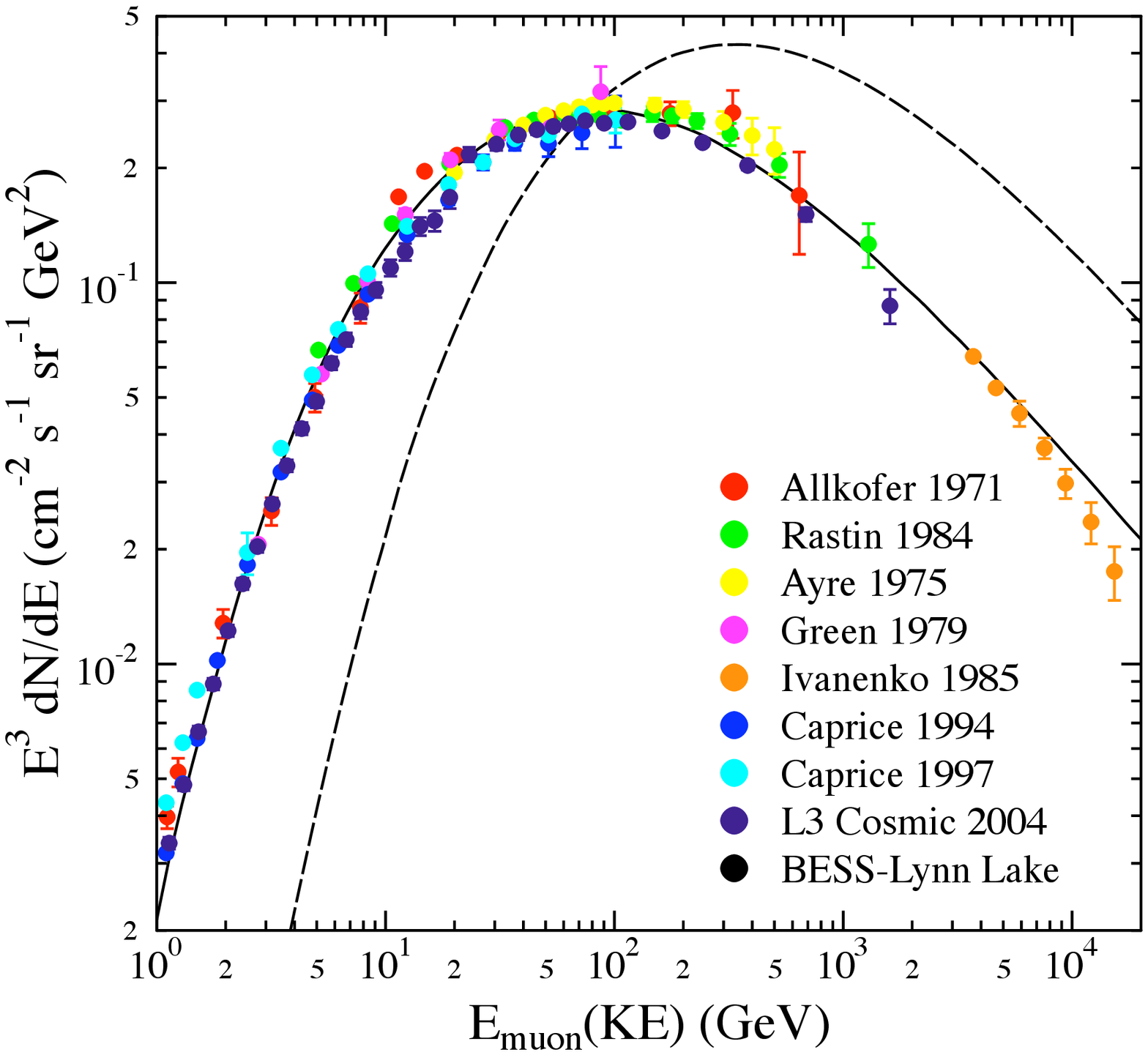}
\caption{The vertical muon flux at the earth's surface calculated from Gaisser's
semi-analytic approach (solid line).  For details and references to data, see Ref. \cite{Gaisser1}.
The dashed line is the flux at 75$^\mathrm{o}$.}
\end{figure}

The calculations were done by two of us (WH and KP), and will be presented
in detail elsewhere (including applications to a several existing laboratories and proposed
laboratory sites) \cite{Philpott}.   The approach follows the work of Gaisser \cite{Gaisser1},
combining his semi-analytic treatment of atmospheric muon production (which Gaisser 
calibrated against cascade-code calculations) with his treatment of muon energy loss \cite{Gaisser2}  in
rock.  Figure 4 compares the predicted atmospheric
flux derived from Gaisser's semi-analytic formula to surface muon flux measurements,
for vertical muons and muons at large angles.  Figure 5 shows the resulting predictions
of the total muon flux underground, compared to measurements from MACRO and LVD 
at Gran Gasso, and from a variety of other underground laboratories (WIPP, Kamioka,
Boulby, etc.)   Also shown is the comparison with the parameterization of
Mei and Hime \cite{hime}.  The agreement is very good.  In the energy-loss formula
\begin{equation}
- {d E_\mu \over dx} = \alpha + {E_\mu \over \beta}
\end{equation}
where $x$ is the depth in standard rock in units of km.w.e., we determined $\alpha$ = 
200 GeV/km.w.e and $\beta$ = 2.5 km.w.e.
from the data on muon fluxes in deep laboratories, yielding the fit shown in
Fig. 5.  These values are quite consistent with the relevant high-energy values recommended
by Gaisser, e.g., $\alpha$ = 268 (293) GeV/km.w.e. and $\beta$ = 2.55 (2.30) km.w.e.
at 1 TeV (10 TeV) \cite{Gaisser2}.

\begin{figure}
\includegraphics[width=9cm]{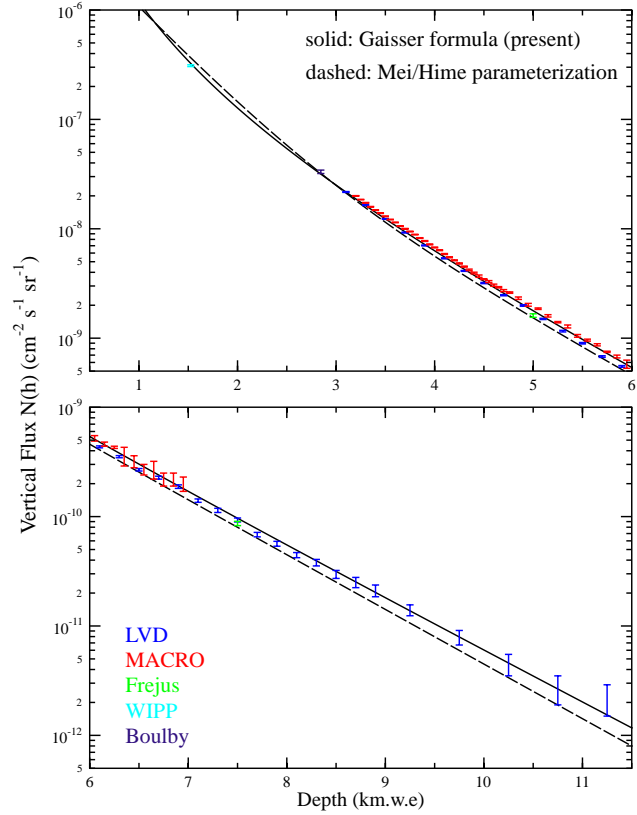}
\caption{The muon flux as a function of depth in standard rock.  The solid line gives
the results computed from the surface muon flux and energy loss formulas of
Refs. \cite{Gaisser1,Gaisser2}, as discussed in the text.  The dashed line is the
empirical fit to underground measurements of Mei and Hime \cite{hime}.  For
references to the data see \cite{Philpott}.}
\end{figure}

\begin{figure*}
\includegraphics[width=17.5cm]{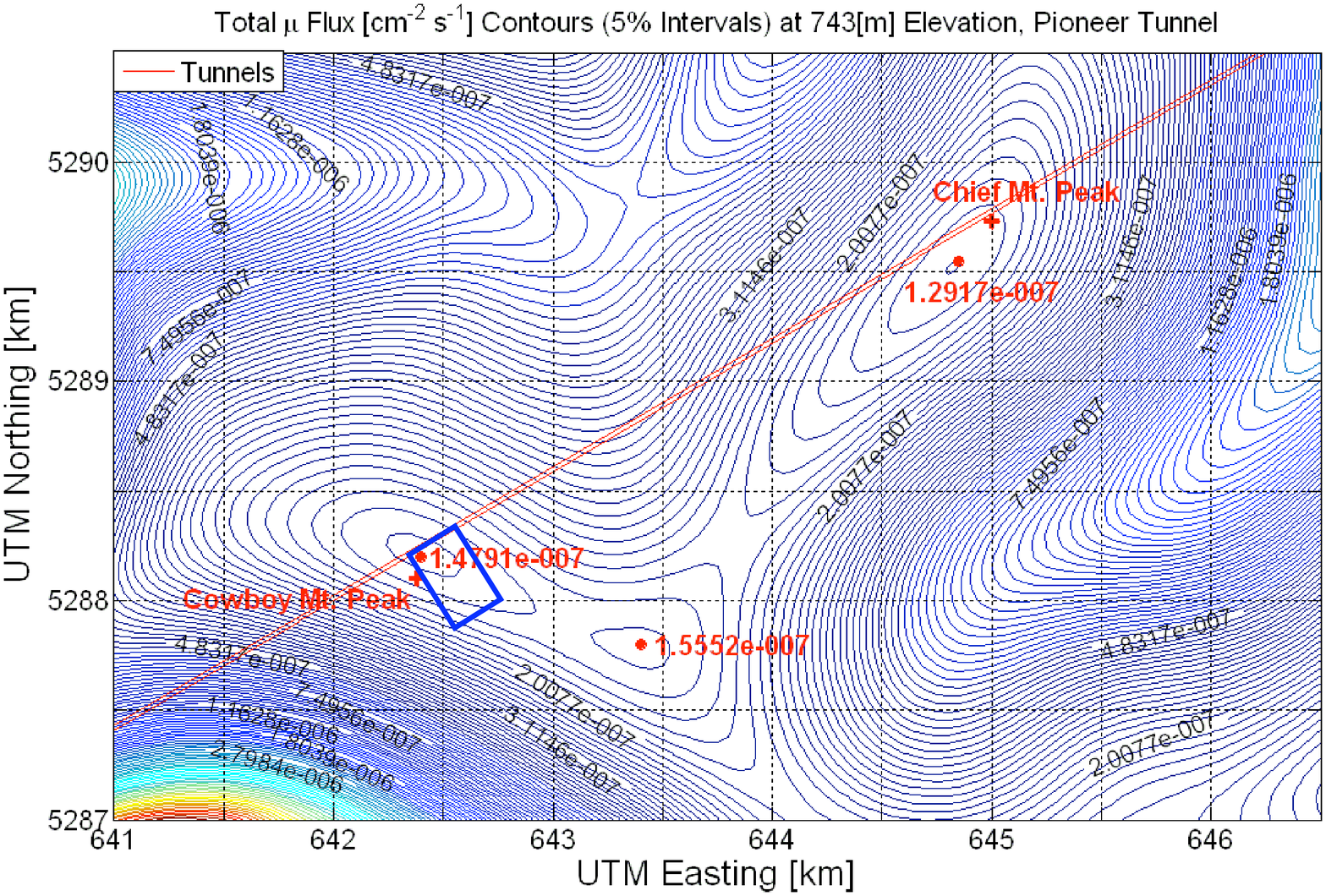}
\caption{The cosmic ray depth contour map for the Stevens Pass area in the Mt. Stuart
batholith, evaluated at 743m above sea level, the elevation of the Pioneer tunnel
as it passes under Cowboy Mt.  The tunnel alignment is ideal for a
Stage I laboratory developed immediately off the south wall of the Pioneer tunnel
(area enclosed in the rectangle), as
shown in Fig. 9.  The contours correspond to 5\% changes in effective overburden.
There is an extended ridge south and east of the Stage I site where one could
locate a very large detector (perhaps at the end of a downgrade hallway extending from
the Stage I area, as indicated in Fig. 9).  
The cosmic ray flux at the Stage I site is 
1.48 $\times$ $10^{-7}$/cm$^2$ s.}
\end{figure*}

One result obtained from the DEMs and satellite photos of the portals is an accurate
tunnel alignment.  The position of the tunnels in the
report of Ref. \cite{SW2}, taken from US Forest Service maps which place Cowboy Mt.
and Big Chief Mt. slightly north of the tunnels, proved to be somewhat in error.  
The peak of Cowboy Mt. is approximately 140m south of the Pioneer tunnel; and region of
maximum overburden is approximately 60 m south of the tunnel.   Thus the tunnel
position is ideal from the perspective of science: rooms developed off the south wall
of the Pioneer tunnel, as required by BNSF to keep new excavation away from the
main tunnel, would be at the point of maximum overburden (see Fig 1).

\begin{figure*}
\includegraphics[width=17.5cm]{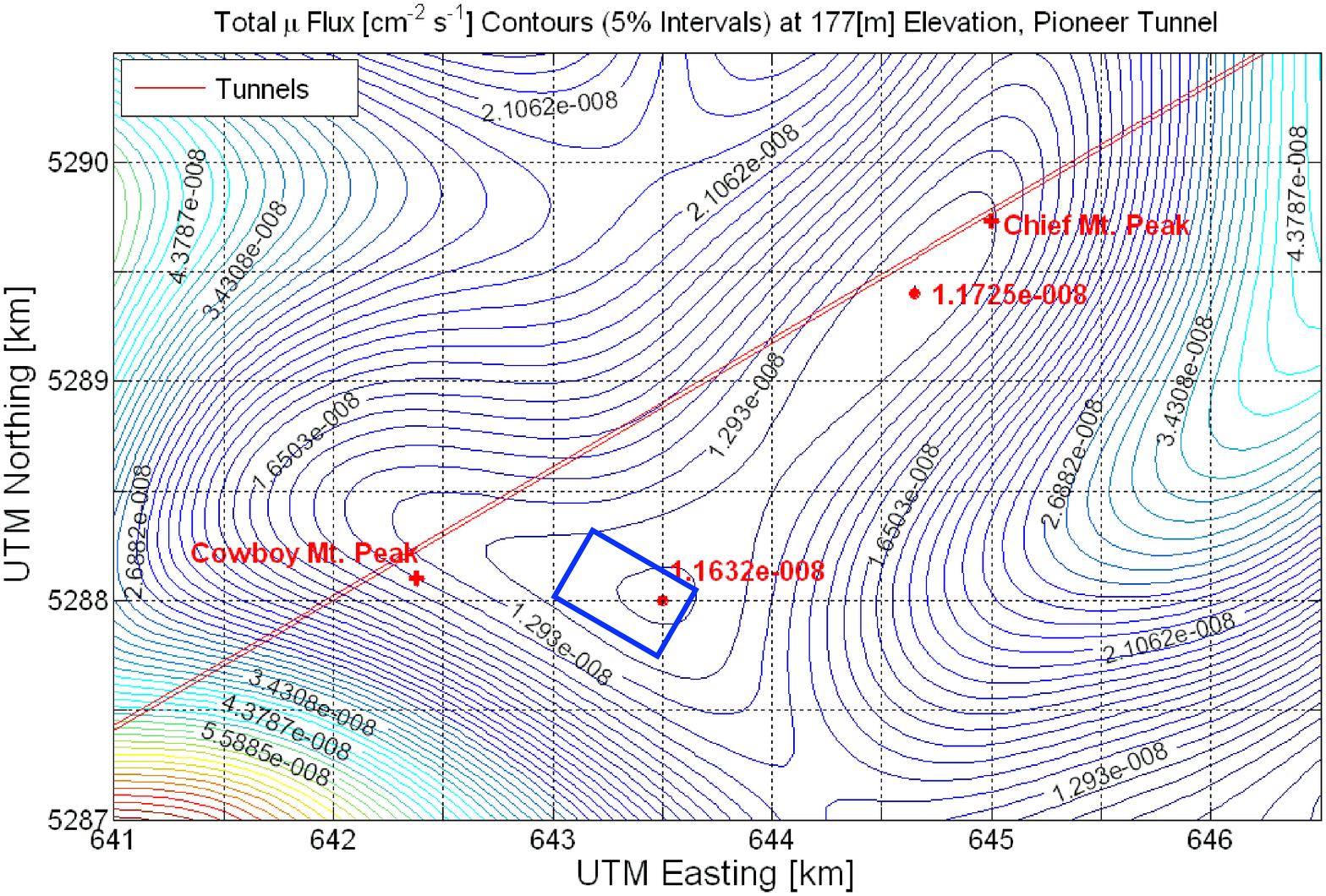}
\caption{As in Fig. 6, but evaluated at the Stage II evelvation of 177m.
The minimum cosmic ray flux at the Cowboy Mt. Stage II site (enclosed in the rectangle) is 
1.16 $\times$ $10^{-8}$/cm$^2$ s.}
\end{figure*}

Figure 6 is the ``muon depth" contour plot appropriate for Stage I, calculated for  
an elevation of 743m above sea level,
the elevation of the Pioneer tunnel at Cowboy Mt.   The contours correspond to changes in
muon flux of 5\%.  The
flux at the proposed laboratory location is 1.48 $\times$ 10$^{-7}$/cm$^2$ s.  This corresponds to the muon
flux that would exist below a flat surface at a depth of 2.12 km.w.e. (all calculations
assume standard rock).  
To calibrate this depth
against existing laboratories, the same code was used to integrate the muon fluxes for
Kamioka and Gran Sasso, yielding 1.75 $\times$ 10$^{-7}$ and 2.96 $\times$ 10$^{-8}$, respectively,
corresponding to depths of  2.04 km.w.e.
and 3.03 km.w.e.   This shows that the Pioneer site is just slightly deeper than Kamioka.

No attempt was made in these calculations to account for density differences 
between Chiwakum schist and Mt. Stuart granodiorite or within the batholith, or for
deviations in rock chemistry from that of standard rock.  These
refinements may be considered in future work.  We anticipate that the effects of density
variations will be small, as the densities found in field studies were confined to
a narrow range.

Figure 6 shows that an area near Big Chief Mt., 2.8 km to the east, provides 10-15\% 
greater attenuation.  However, the elevation of the Pioneer tunnel at Big Chief Mt. is about
45m above that at Cowboy Mt., removing most of the difference.  Cowboy Mt. is the
optimal location because it is closer to the Scenic portal, reducing travel time to the surface
and the length of tunnel that would be improved (see Section III).  

Another feature of Fig. 6 is the ridge that forms south and east of Cowboy Mt., an extended
region of high overburden.  If any large cavity were excavated at the Stage I site, it would
be placed on this ridge, well away from the tunnels and below tunnel grade.  The site
has the potential to provide more than adequate separation with no loss in overburden, as
discussed in Section III.

In Section V we will discuss a possible Stage II development at the Pioneer tunnel that could be
undertaken in parallel with Stage I or at any time thereafter, depending on the point
when the U.S. science community finds itself in need of additional underground space
at greater depth.  The proposed Stage II would be located 566m below Stage I, at an elevation under 
Cowboy Mt. of 177m above sea level.  Figure 7 shows the flux contours at this elevation.
There is an extended region along the ridge where
the overburden is quite uniform, corresponding to a flux reduction of $\sim$ 13
from Stage I levels.   This would be helpful in positioning Stage II so that it can make use
of the ventilation and rail access available on the Stage I level.
The muon flux at the Stage II location would be 1.16 $\times$ 10$^{-8}$/cm$^2$s, equivalent
in a flat site to a depth of 3.62 km.w.e., and about 2.6 times lower than that of Gran Sasso.
Opening this space requires 
approximate 4.8 km of tunneling, as described in Section V. 

\section{Stage I Potential}
In this section we will outline the improvements we believe would optimize a Stage I laboratory (the
Kamioka-depth stage).   The site offers existing 
horizontal access that can be dedicated to science, which is very unusual.  It also has
many of the qualities of a greenfield, that is,  a purpose-build site, despite the pre-existing 
access: the tunnel has
never been used by the railroad, except as a gallery for groundwater drainage.  The 
site provides
a dedicated entrance tunnel, a separate ventilation/utility tunnel,  and crosscuts to a
refuge tunnel with an independent ventilation system,
elements one would design in much the same way in a purpose-built site,
were the resources available.  The site has unusually stable redundant power and
excellent highway and railway access.   

Measuring distances from the Scenic portal, we will denote the portion of the tunnel between 0
and 3.7 km as the entrance tunnel, between 3.7 and 3.95 km as the laboratory, and between 3.95 and
8.6 km as the ventilation/utility tunnel.   Laboratory construction would begin by enlarging the
entrance and laboratory sections of the tunnel
from the current 9 ft by 8 ft profile, using smooth-wall blasting techniques.   Fig. 8 shows the desired
profile (approximately 15 ft (4.6m) in width, and arched to 15 ft).  This would provide an 11.5 ft. roadway,
the same width as was used in the recent rehabilitation of the Whittier tunnel in Alaska (a project
with a number of similarities to the one proposed here) \cite{whittier}, as well as an adequate
walkway for maintenance and emergency egress, 0.6 m in width.   The minimum overhead
clearance is also approximately 11.5 ft.  Such tunnel rehabilitation and enlargement is
very common: the usual procedure is to raise the crown of the tunnel, and then to
finish the walls and crown with steel-fiber-reinforced microsilica shotcrete \cite{harvey}.
Bolting and other additional support requirements would be addressed, as discussed in \cite{SW2},
though the needs should be minimal.

\begin{figure}
\includegraphics[width=9cm]{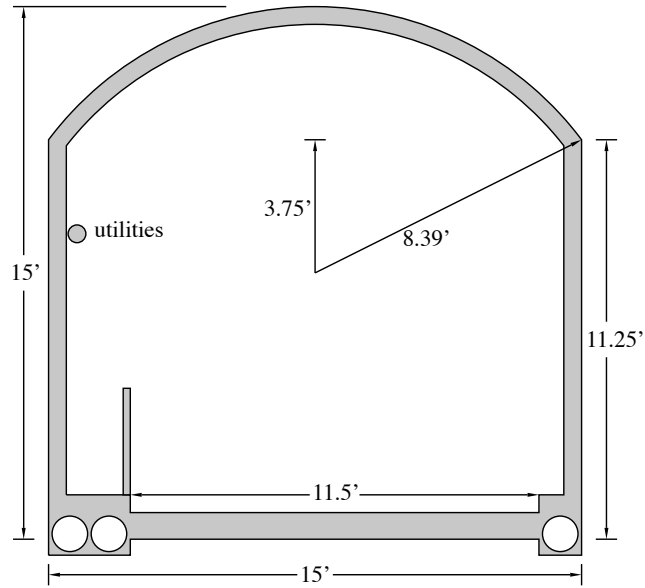}
\caption{The schematic of the proposed profile of the Stage I entrance tunnel, after enlargement of
the Pioneer tunnel,
shotcreting, and concrete floors.}
\end{figure}

The floor would be tracked but drivable, with the tracks connected at the portal to BNSF
operations, making the laboratory a siding off the railway.  Rail access simplifies
excavation.  Rock produced in tunnel enlargement or
in laboratory excavation would be loaded underground onto rail
cars and, if not wanted by BNSF, hauled to one of the nearby commercial aggregate
pits.  (Several large
aggregate companies have located their facilities near the railway, west of Scenic.)

The tunnel's 1.56\% slope allows gravity drains to work well.   For example, each of the 
three 12-inch-diameter pipes shown in Fig. 8 would have a capacity of about
2000 gpm, well above the total current drainage from both tunnels of about 600 gpm.
Thus such a configuration would provide substantial excess capacity that could be
put to use in Stage II, for example.    The water velocity 
in a 12-inch pipe inclined at 1.56\% is approximately 6.2 ft/sec, 
well above the 1.5 ft/s that would allow sedimentation to occur.   The north-wall drain pipe would
be attached to the drains from the main tunnel, with sealed connections.
The two south-wall pipes would allow one to keep Stage I and Stage II drainage
sequestered, which could be helpful in environmental monitoring, while also
separating all Pioneer tunnel and laboratory drainage from main-tunnel drainage.
The drainage system could be placed
under removable gratings, to simplify inspections and maintenance.
 
The remainder of the tunnel (that is, the ventilation/utility section to the east)
would be refurbished and resupported, but not enlarged apart from
minimal microblasting designed to smooth the rock walls.  It is likely that a tunnel cross section
of about 100 ft$^2$ would result, sufficient to handle air flows of up to 180,000 cfm in emergency
mode (far beyond any requirement for Stage I).  We would
advocate extending the concrete floor, track, and a portion of the drain-pipe system to the 
Pioneer tunnel's eastern terminus at Mill Valley.  This
would simplify construction, make the unlined ventilation tunnel more convenient  for possible
earth science studies, and contribute to safety in the main tunnel, as noted below.
In addition, existing cross cuts would be improved and enlarged.

The proposed ventilation intake is Mill Valley, a forested area within the BNSF property, located far from
railroad operations and Highway 2.  The air quality in the valley is typically very good.
While the existing shaft at Mill Valley could be reopened, it might be less costly to reconnect 
the Pioneer tunnel to the surface via a new, small-diameter ventilation bore.   
A 2.6m-diameter concrete-lined raised-bore 
ventilation shaft could be constructed quickly and very economically.  The drill cuttings would
be carried away using the Pioneer tunnel tracks.  Such a shaft could carry 
180,000 cfm at an economic velocity of about 3000 ft/m.  Airflow in the Pioneer tunnel would be
east to west, which would help keep diesel emissions (produced by delivery trucks
arriving at the laboratory) away from the laboratory area.  The BNSF's main-tunnel ventilation system also pushes air from east to west.

The design of the ventilation system would depend on the types of experiments housed in the
facility.  A simple flow-through ventilation system is employed in other rail-tunnel laboratories, such
as the Oto Cosmo Laboratory and Canfranc, but safety considerations then preclude experiments utilizing large volumes of cryogens or flammables.  However, there are
many options for implementing more sophisticated systems, including designs similar to
that described in \cite{DC}.  In that plan a system of ducts isolates air flowing
to laboratory areas from tunnel air flow, recirculation is used to achieve cleaner lab spaces, and a system of vents and reversible fans allows a tunnel fire to be isolated and exhausted.

One design of intermediate complexity combines the simple east-to-west tunnel flow with a 
second, separated ventilation system that brings clean air from the surface to laboratory
areas through an overhead duct.  Such a system could also lower laboratory radon levels.
There is a second convenient access for a clean air intake, the Tye Incline, a point of
low overburden about 650m from the west portal.  The ducted air,
when exhausted from the laboratory, would then join the tunnel airstream.
As in \cite{DC}, a second duct for ``special processes" exhaust from the
laboratory could run to the west portal, to keep industrial air away from occupied tunnels.
Thus this system would require two ducts to be installed along the ceiling of the
entrance tunnel.  Finally, the fans controlling tunnel ventilation would be made
reversible.  This would allow an emergency ventilation mode, in case of a fire in a
detector, in which laboratory air is exhausted to the east, through the ventilation/utility tunnel, while
personnel evacuate to the west, through the entrance tunnel and against the (now
reversed) air flow.

Stevens Pass is a major power and communications corridor.  The transmission line from Rocky Reach Dam on the Columbia River to Seattle follows Mill Valley, passing within 100m of the old shaft.
We would propose bringing the laboratory's main power underground from this point.  Mill Valley
is in Chelan County, where power costs are exceptional low due to the utility district's
rights to Columbia River power.  A second major
transmission line, from Chief Joseph Dam, runs by the Scenic portal.  Thus power from this source
could be brought to the laboratory through the entrance tunnel, as a backup supply.

\subsection{Laboratory development} A conceptual sketch of a possible laboratory layout
is shown in Fig. 9.   The design provides drive-in access to the experimental area, a SNOLab-like
clean laboratory with a car wash/personnel entrance to the west, three experimental rooms,
and ventilation flow from Mill Valley, down the ventilation/utility eastern portion of the Pioneer 
Tunnel.   The two crosscuts shown in Fig. 9 
are denoted 8a and 9 in the Shannon \& Wilson
report \cite{SW1}: these can be used as markers to superimpose Fig. 9 on the Pioneer
tunnel geology, support, and hydrology maps contained in that report.

\begin{figure*}
\includegraphics[width=16cm]{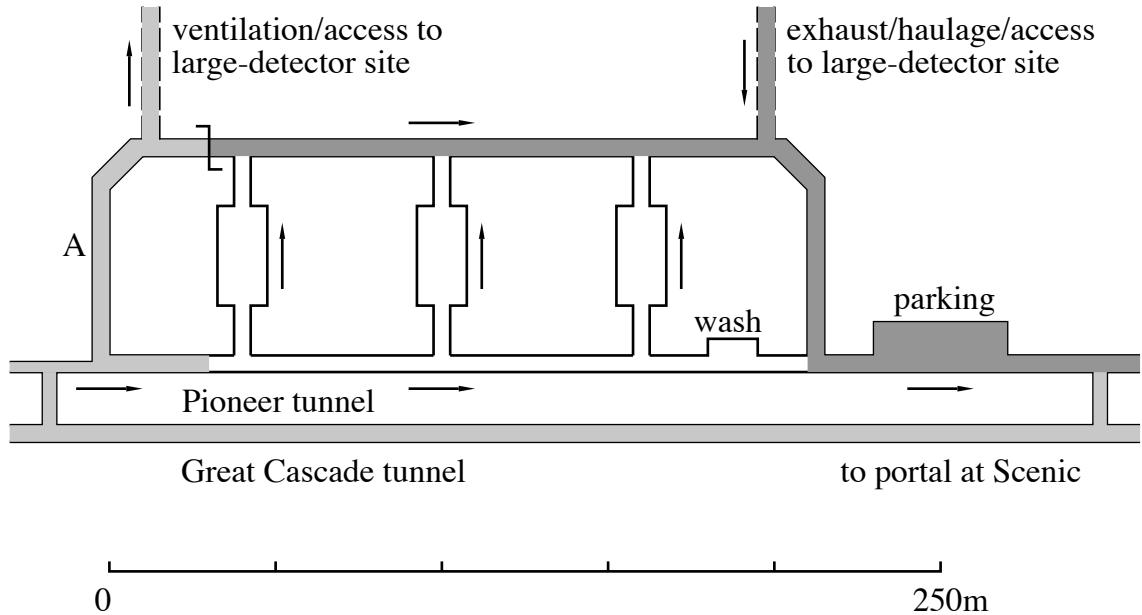}
\caption{A schematic of the envisioned Stage I laboratory in which the eastern portion
of the Pioneer tunnel becomes a ventilation/utility hallway and the western portion
becomes a dedicated entrance hallway.  Rooms are spaced at three times their
spans.  The ventilation flow is indicated, with lightly
shaded areas indicating incoming air, darkly shaded areas exhaust, and unshaded areas
laboratory space behind the clean barriers.  The resulting ``dirty-side" access would
allow the laboratory to be developed further without impacting ongoing experiments, e.g.,
as indicated for a possible large-detector development.  The crosscuts shown are 8A
and 9 on the Shannon \& Wilson geology/tunnel support/hydrology maps, which allows
one to superimpose this plan on the known geology \cite{SW1}. }
\end{figure*}

The design provides every room with two exits, and allows exiting personnel to leave either
exit in either direction, with one direction therefore always being against the air flow.   Clean
laboratory areas -- areas behind the clean barriers served with filtered air -- are unshaded.  
Air intake paths are lightly shaded, and exhaust paths are more heavily shaded.  For
the purpose of illustrating a possible room configuration, three  
15m by 30m rooms suitable for midscale experiments, such as double beta decay, dark
matter, low-level counting, or nuclear astrophysics, are shown.   A conservative spacing
between openings of three times the room span is maintained.  Rooms are set back from 
hallways by 15m.

The Pioneer tunnel track would allow loads to be brought in on a railed car to and through
the carwash, directly to laboratory hall entrances.   As proposed in \cite{DC}, lab personnel
could drive from the portal to the laboratory using electric vehicles similar to those employed at airports.
A parking area is provided for this purpose. 

The ventilation requirements for a laboratory so configured can be estimated.  Combining the
total laboratory (room and hallway)  floor space of 2600 m$^2$ with 
an exhaust requirement of 0.75 cfm per square foot of floor space yields a 
normal operations flow of 21,000 cfm.  In the case of an emergency in a laboratory room,
assuming 8m ceilings, a flow of 13,000 cfm through that room would be sufficient to produce
six changes of air per hour.   Thus a ventilation system designed for 25,000 cfm would easily meet
both normal operations and room emergency requirements.  This corresponds to a modest air
velocity requirement in the ventilation/utility tunnel of 250 feet per minute.  Assuming relatively 
smooth but
unlined walls in the ventilation/utility tunnel and and a tunnel cross section of 100 ft$^2$, the 
necessary fan pressure is found to be a minimal 0.3 inches of water gauge.   However, the requirement
for emergency ventilation of the entrance tunnel, assumed to have a cross section of
about 200 ft$^2$, places greater demands on the fan.  A sustained emergency 
flow of 500 ft/min (or 100,000 cfm) in the entrance tunnel (adequate to prevent
backlayering of smoke) requires a pressure of about 4.8
inches of water gauge,  fan power of about 115 horses, and an air velocity
of about 1000 ft/min in the ventilation tunnel.  These specifications are still well within the
capabilities of a standard one-stage system.  One concludes than normal-operations 
ventilation and power requirements are minimal, and that emergency operations for entrance
tunnel evacuation could
be addressed by a single-stage variable fan with such emergency capabilities.

While a more complicated system could be designed (e.g., one with occasional turnouts 
along the entrance tunnel), a laboratory of this scale could easily operate with the restriction
of one-way traffic flow in the entrance tunnel.  A signal system could indicate whether the 
tunnel is open for either entering or exiting traffic.  

The laboratory configuration of Fig. 9 can be further developed without jeopardizing
the operations or cleanliness of experiments.   Additional rooms can be added on the
east end successively, with all excavation done via the exhaust tunnel, and with 
clean ventilation provided to the work air via the pathway marked A.   In the same way, large-detector
construction could be carried out at an appropriate distance from the main laboratory:
such a detector could be placed south of the laboratory, along the ridge shown in 
Fig. 6.  Most likely one would want the detector cavity to be below the level of the main
laboratory.  In this case, the ventilation/access and exhaust/haulage/access tunnels
would be declined.  A convenient excavation plan would be to extend track along the
laboratory's exhaust tunnel, so that cars would be moved to this location.  Then crushed
rock from the excavation could be brought to the exhaust tunnel by a short conveyor
running along the tunnel marked exhaust/haulage/access
in Fig. 9, and loaded directly onto rail cars.
The route along the Pioneer tunnel and to nearby pits is downhill.  The market value of
the crush rock that would be produced in a project like UNO (Ultra underground Nucleon
decay and Neutrino Observatory \cite{jung}) is about \$20M. 

\subsection{Development plans and costs}
Here we briefly outline the steps that we would recommend to bring the Pioneer tunnel
facility online and to operate the facility with 24/7 access.  An advantage of this site
is that it can be developed rather quickly, operated at exceptionally low expense,
and then later expanded in successive stages to achieve greater depth, as needs
arise.  This allows the laboratory to build an experienced staff and user group, then
utilize this expertise in developing a sensible plan for future stages.

The estimate of excavation and finishing costs
(bolting, shotcrete, mineguard, concrete floors, drainage, etc.) of \$27M is based on Ref. \cite{SW2}.
Approximately \$6.5M of this total is connected with engineering 
choices that would be of primary benefit to main-tunnel users, enhancing its
safety, drainage, and accessibility.  We discuss in the next subsection some of the
potential partners who might benefit from such improvements.

$\bullet$ Entrance tunnel:  The principal construction task will be the improvement of
the access tunnel.  The first 3.95 km of the Pioneer tunnel would be enlarged to the
profile shown in Fig. 8, corresponding to an opening 15 ft in width and peaked
to 15 ft, before shotcreting and concrete work.  In Ref. \cite{SW1} it was noted
that similar tunnel enlargement projects have led to excavation costs of between
\$50-90/yd$^3$, depending on access and rock quality issues.  The value recommended
in Ref. \cite{SW1} for the purpose of Pioneer tunnel estimates is \$63/yd$^3$.  
To achieve the profile of Fig. 8 approximately 5.4 yd$^3$/ft would need to be excavated.
We would also proposing enlarging and improving the nine crosscuts in this part
of the tunnel.  The estimated excavation total would be approximately  \$4.5M 
(3.95 km of entrance tunnel and laboratory hallway).

While some plans for deep facilities (DUSEL) envision access through unlined drifts,  there
are a variety of cleanliness, safety, and maintenance reasons for lining tunnels, despite
the additional initial cost.  We would propose lining the entrance tunnel and laboratory
hallway walls and ceiling with shotcrete.  The costing in Ref. \cite{SW1}
also provides for some bolting to provide additional support, at \$40/ft.  The total for 
tunnel lining and support is \$5.9M.  Finally, the cost of 9-in concrete floors, subfloor
drainage, and track is estimated to be \$1.6M.  Thus total entrance tunnel development
costs -- excavation, lining, floors, drainage, track, and cross cut improvements -- 
would be approximately \$12M.

$\bullet$ Ventilation/utility tunnel:  While rehabilitation of the ventilation/utility 
tunnel could in principle be avoided by ducting air into the laboratory, this would eliminate
many possibilities for ventilation upgrades that we feel would be important to some 
Stage I experiments.  An upgraded ventilation tunnel is also important to earth science,
providing access to 4.7 km of unlined tunnel and a convenient location for
installing long-term observatories, such as boreholes instrumented from the surface to the
tunnel.

The ventilation/utility tunnel rehabilitation would have a number of positive impacts
beyond the project, including the opportunity to increase the safety of and access
to the main tunnel, which is used by both freight and passenger trains.  These issues
are discussed below. Thus we would expect to find partners to help us with the cost.
The proposed work includes smoothing the tunnel walls to provide a
minimum cross section of 100 ft$^2$ (\$1.0M), resupporting the unlined tunnel
according to the prescription of Ref. \cite{SW1} (\$150/ft, for a total of \$2.3M),
installing concrete floors, track,  and underfloor drainage (\$1.5M), timber set and debris
removal (\$0.85M), improving 9 additional cross cuts (\$0.1M), and boring a 660-ft
2.6-m concrete-lined ventilation shaft  (\$0.7M).  The total is approximately \$6.5M.  In
return for this investment, a modern access/escape tunnel with a separate atmosphere would be
available along 8.6 km of the Great Cascade tunnel.

$\bullet$ Room and laboratory hallway development:  The excavation, support, and
finishing costs for rooms of the dimension discussed here are available in
Ref. \cite{DC}.   The estimates include shotcrete, mineguard coating to reduce radon,
a concrete floor, track in the room entrances, and bolting for the walls and ceiling.
The total for three rooms is \$1.85M.  The excavation and similar finishing of approximately 450m of 
connecting hallway and exhaust tunnels would require an additional \$1.8M.  
Finally, the excavation and finishing of the car wash and parking area, again
based on the estimates of Ref. \cite{DC}, requires about \$0.54M.  Thus the total for interior
laboratory construction and finishing is \$4.2M. 

$\bullet$ Additional costs:  Mechanical systems that would need to be
designed and costed for such a laboratory include ventilation and air filtration;
a water storage tank, water heater, and piping; and fire protection.  Electrical needs
include switching gear and unit substations, 480V distribution, and lighting.  These are
not costed here.  Finished laboratory rooms generally require a substantial additional
investment to provide the utilities necessary for clean operations.  Challenging
experiments, such as a those using large volumes of cryogens, may require pressure
doors and a ventilation system meeting stricter safety requirements, as well as 
pits placed below level.

$\bullet$ Non-scientific operations economies in a horizontal-access
laboratory:  It is generally understood that operations costs of horizontal
access facilities can be many times smaller than costs of facilities requiring hoist
access.  For example, the manpower required to maintain 24/7 deep mine access 
in the study of Ref. \cite{Homestake} was determined to be between 58-84 FTEs, 
depending on the investment in hoist modernization to reduce operator requirements.
Operations can dominate the project lifetime costs in such facilities.

The manpower requirements to maintain 24/7 access to Pioneer tunnel facility 
needs would be quite modest.   A watchman would be needed at the portal entrance
at all times and two or three staff would be needed to maintain the ventilation, air
filtration,  and drainage
systems and monitor environmental and safety controls.   The facility power
needs are exceptionally low, the drainage system is gravity powered, and the utility
rates in the host county are among the nation's lowest.

Note that the discussion above is restricted to non-scientific operations.  Additional personnel are needed for scientific operations: technicians to maintain cleanrooms
and specialized experimental utilities, mount detectors, handle refrigeration and electrical
needs of experiments, etc.  Some of these scientific personnel requirements would be similar in
horizontal and vertical laboratories.  But others, such as the support the laboratory has
to extend to experimenters bringing detector components underground, would be 
reduced in horizontal facilities because of the ease of access.

\subsection{Broader impacts of Pioneer tunnel improvements} 
One attractive aspect of this project is that its execution will address a number of 
concerns important to BNSF, Amtrak, and state and federal agencies concerned
with economic growth and public safety.\\

\noindent
\textbf{\textit{Safety Issues:}}  In modern tunnels, such as the Chunnel, it is recognized that an
independent service/rescue tunnel greatly enhances safety and simplifies maintenance.   One of
the reasons we propose using the eastern 5 km of the Pioneer as a ventilation intake --
including finished floors with underfloor drainage -- was the recognition that this design
would convert the Pioneer tunnel into a high-quality service/rescue tunnel, at relatively
modest cost.  The work includes
enlarging and improving the existing crosscuts, so that they would be adequate 
for the general public, such as passengers using Amtrak.   The laboratory ventilation
system does double duty, providing clean air to the laboratory while maintaining an
atmosphere in the rescue tunnel that is separate from that of the main tunnel.  Our plan to bring
power from Mill Valley enables us to place emergency lighting in the ventilation
tunnel.  The concrete, railed flooring would allow rescue vehicles to reach the 
cross cuts to evacuate personnel, in the advent of a main-tunnel emergency.

This plan is somewhat more costly than the alternative, 
bringing ducted ventilation up the entrance tunnel from
the Tye Incline, thereby eliminating any need to refurbish the eastern 5 km
of the Pioneer tunnel.  However, the additional investment not only contributes to
public safety, but also pays some scientific
dividends, as noted previously.   A separate ventilation tunnel
provides additional flexibility in future ventilation schemes, should a Stage II be
pursued with a ventilation system coordinated with that of Stage I.

We note that the Great Cascade tunnel has featured prominently in Congressional
discussions of public safety and homeland security \cite{MSNBC}: it is recognized as a point of
vulnerability on one of the nation's most important economic lifelines, as 
public safety challenge due to the difficulty of evacuating a tunnel of such exceptional
length, and as a strategic asset because of its role
in rail support of Fort Lewis (the only Power Projection
Platform -- or mobilization center -- on the west coast).

\noindent
\textbf{\textit{Rail Capacity:}} The BNSF Northern Route is one of the nation's most important
freight lines, linking the Puget Sound ports of Seattle and Tacoma with Chicago
and the Midwest.   These ports now rank second nationally in the container traffic
they handle, which exceeded 4 million TEUs (twenty-foot equivalent units) last
year.  The growth rate, driven by Pacific Rim and NAFTA trade, has been 15-20\%/year
in recent years.  Approximately 70\% of the container traffic is moved eastward, with
the Great Cascade tunnel being the preferred route, as it is the only passage through
the Cascades with sufficient clearance for double-stack container cars.  

The route is now nearly saturated at its capacity of 25-30 trains/day.  One bottleneck
is the Steven Pass area, due to the single-track in and around the Great Cascade
tunnel and the ventilation requirements of that tunnel.  During the 30 minutes an
eastbound (upgrade) train is in the tunnel, the ventilation system must force
air around the engine, to cool the engine and to provide adequate oxygen for combustion.
In the case of consecutive eastbound trains, after the first train
has exited at Berne, the fans must operate at full capacity for
an additional 30 minutes to force
exhaust trapped in the empty tunnel out the portal at Scenic, before
the second train can enter.  That is, the minimum separation between consecutive
eastbound trains is about one hour.   The development of the Pioneer
tunnel laboratory would have some positive impacts on tunnel capacity:\\
$\bullet$  There is a loss in efficiency in cooling eastbound trains due to leakage of
air into the Pioneer tunnel and out the Tye Incline, as this opening was not adequately sealed
when closed.
This leakage reduces the amount of air that the main tunnel
ventilation system can force around eastbound trains, limiting the cooling.
Consequently the engines of some eastbound trains overheat, forcing the engineers
to reduce speed to 8-10 mph, typically.  At this speed the tunnel transit time is doubled
to approximately an hour.  This may occur once or twice a day.  The 
laboratory developments we described would seal
the drains from the Great Cascade tunnel (the path by which air leaks into the
Pioneer tunnel) and the Tye Incline (which is located in an area of the tunnel to be
shotcreted).  This should completely isolate the Pioneer tunnel from main-tunnel
ventilation, so that air leakage and associated overheating are eliminated, potentially
increasing tunnel capacity by 5-10\%.  \\
$\bullet$  The blockage of drains has been identified by Shannon \& Wilson as a
potential maintenance issue for the main tunnel.  The Pioneer tunnel upgrade would
clear all debris and produce a superior, easily maintained drainage system.\\
$\bullet$  The rate of increase in rail haulage will soon force some major improvement in
the capacity of either the Stevens Pass or Stampede Pass routes.  (Stampede Pass,
the other mainline route through the Cascades, has a tunnel that lacks adequate clearance
for double-stack cars.).  An issue in any future construction project at Stevens Pass 
will be interference with existing traffic in the Great Cascade tunnel.  The Pioneer tunnel
improvements we have proposed for science would also restore the tunnel to its
original purpose, alternative construction access to the main tunnel.
Construction crews would be able to reach the midpoint of the Great Cascade tunnel
via the Pioneer tunnel, using rubber-tired or railed vehicles, without interfering with either
laboratory operations or main-tunnel activities.  There are projects that could be
completed with such access, to improve main-tunnel capacity.  One we have explored \cite{SW2}
is the installation of a two-zone ventilation system like that employed in Canada's
MacDonald Tunnel (North America's longest).  This might be done, for example, by boring
ventilation shafts near the main-tunnel midpoint: the cuttings produced
could be removed via the Pioneer tunnel.  Such a two-zone ventilation system would allow
more efficient clearing of the main tunnel, reducing the minimum
separation between consecutive eastbound trains from the present 60 to about 35 minutes.
There are other conceivable tunnel improvements, such as creating a midtunnel siding, 
that could be effectively supported from the Pioneer tunnel.\\

\noindent
\textbf{\textit{Monitoring and Interrogating Shipping Containers:}}  One post-9/11 security concern is the deployment of instrumentation
for efficiently interrogating the vast volume of material brought onshore in sealed shipping
containers.   The Great Cascade and Pioneer tunnels could play an important role in this effort.
The Cascade tunnel is one of the major choke-points for 
container traffic nationally: a significant fraction of the TEUs (twenty foot equivalents) brought
onshore passes through this tunnel.  Its unique feature, in the national
transportation grid, is the extended region of low background it provides, in principle 
allowing moving containers to be interrogated for up to 30 minutes. If the Pioneer tunnel is developed
for science the way we propose, this site would also provide the low-level counting 
expertise, computing and networking, and physical access necessary to conduct a sophisticated 
program of container interrogation, without interfering in any way with normal transportation
through the main tunnel.

There are several gamma-ray signals expected in nuclear materials, including at least
one in highly enriched
uranium, that might be more easily detected deep underground, given large-area
detectors with good resolution.  The cross cuts would be the natural locations for detector arrays,
as the detectors could then be maintained and monitored from the Pioneer tunnel,
without interfering with main-tunnel activities.  An investment
in a large-area detector would be appropriate given the site's exceptional container traffic
and isolation from the public.

Homeland Security concerns are prompting improved
methods of tracking and identifying cargo containers, such as
optical scans and radio frequency identification.
It might be possible to combine container identification with
sophisticated, low-background detection systems to build large databases.
The Great Cascade tunnel would be a interesting location for such an effort because
of the container traffic volume and the opportunities described above for developing and
maintaining unusually sensitive detectors. 
Ideally the system would be automated so that, as trains pass through the
tunnel, the IDs of each container could be read and recorded.  Properly done, this
system would identify the cargo (according to manifest) while also continuously
monitoring the  train's position within the tunnel.
The manifest would then be correlated with the signatures recorded in
$\gamma$ or neutron detectors that would be distributed at various
locations throughout the tunnel.  Such a database would
be useful in assessing issues such as the envelope of signals that result from similar
containers, the accuracy of manifests, the extent of seasonal variations
in responses due to changing cargo and packing, and the minimum requirements for
effective passive detection systems.

\section{Placing Experiments in SNOLab and the Pioneer Tunnel}
In the abstract, added depth is always good because of the background safety
margin it provides.  However, in the real world background issues are not quite
this simple.  

The first point is that depth is expensive, and thus placing experiments unnecessarily
deep will limit the number of experiments we will be able to do.
The comparative 
construction histories of Super-Kamiokande and SNO are instructive.
Much of the four-year delay in SNO's completion was
directly attributable to the ``ship-in-the-bottle" complexity of construction via hoist and
blue-box.  Indeed, it has been estimated that the more difficult environment of SNO may
have cost the collaboration between 100 and 200 person-years of effort \cite{Art}.  This
depth and thus this cost was necessary in the case of SNO, of course.   But a strategy
that places all experiments deep because a few might need depth is not sensible.
The ``added depth is always good" statement ignores the substantial added costs
of excavation, haulage, and construction deep underground, the associated experimental
delays, and the added manpower requirements.   An experiment like Super-Kamiokande
should be constructed at moderate depth because high-quality horizontal access
is far more essential to the success of the experiment, given a fixed budget, than the
incremental benefits of additional depth (such as reduced deadtime).

{\it One of the important contributions the Pioneer tunnel could make to North American underground 
science is to provide more choices in locating experiments.}  Dedicated horizontal access, rail 
support of major excavation, excellent separation from FermiLab, 
laboratory control of dust and other issues important to background, and a location near
a major center for high-tech industry are among the advantages the Pioneer tunnel offers.
For some experiments these site attributes are very important, but additional
depth is not.

A second point is that depth is often not the most effective strategy for reducing backgrounds,
once radioactivity backgrounds are also considered.  While detectors differ
in their relative responses to cosmic-ray and radioactivity backgrounds, in many
cases these backgrounds becomes about equal at depths of 2.5-3.0 km.w.e.  Thus,
adding depth beyond this point may not significantly reduce overall
background rates.   The more cost-effective strategy
may be to remain at moderate depth but invest in an effective neutron shield that addresses
both cosmogenic and natural radioactivity
backgrounds.  An active neutron shield that is 95\% effective is equivalent
to $\sim$ 2 km.w.e. of additional depth.

A third point is that it is often more sensible to design an experiment for a specific available
site, than to attempt to design a site that will optimize all possible experiments.
For example, low-level counting efforts such as the
double beta decay experiment EXO \cite{EXO} and the dark matter/solar neutrino experiment
XMASS \cite{XMASS}  are being planned for WIPP (1.58 km.w.e.) and Kamioka (2.04 km.w.e.), sites
of moderate depth.   These experiments have properties, such as distinctive signatures
(e.g., detection of the daughter ion Ba in EXO)  or self-shielding detector designs (XMASS),
that reduce sensitivities to backgrounds. 

Indeed, a very vigorous program of next-generation experiments proposed by European
and Japanese scientists will be conducted in
sites like Boulby, Gran Sasso, and Kamioka, despite depths that are
far short of the goal often associated with DUSEL, 6 km.w.e.   In this context, our Pioneer
tunnel proposal is a compromise between the Japanese/European and DUSEL approaches:
we agree with the Japanese and Europeans that much can be done with
shallower sites by  properly designing experiments.  But we also value the Pioneer
site because it can be deepened in an intelligent way, when the time comes that a lack
of deep space begins to impact the science we can do in North America.

These issues are probably best illustrated by examining a list of proposed next-generation experiments.
Table 1 displays a representative set, most of which are in the R\&D phase.  The selection
is quite arbitrary -- some of these proposals may not prove feasible, and other very
strong experiments are not listed.  The list, which is intended as a device for illustrating the types of
background issues that arise in different classes of experiments, reflects  
in part our success in finding quantitative
published analyses of backgrounds.   The table notes the goals,
the shielding assumptions, and the resulting minimum depth requirements, defined in
the table as the point where muon-associated backgrounds and signal are equal,
at the experiment's design sensitivity.   [Note
in the discussion below, we will require such backgrounds to be less that one-third this
value, so that some discovery potential remains at the design sensitivity.]

\begin{table*}
\centering
\caption{Estimates of Required Depth for Representative Future Experiments}
\begin{tabular}{|c||c|c|c|c|c|}
        \hline
Experiment & Goal & Shielding Asssumptions & Min. Depth$^*$ & Ref. & Site \\ \hline \hline
CDMSII DM& 10$^{-8}$ pb $\Rightarrow$ 1 event/kg/y, 10-100 keV & Cu, $\sim$50cm polyeth., 22.5 cm Pb  & 1.2 km.w.e. & \cite{hime} & \\
& 10$^{-9}$ pb $\Rightarrow$ 0.1 event/kg/y & " & 2.6 km.w.e. & \cite{hime} &\\
SuperCDMS& 10$^{-10}$ pb $\Rightarrow$ 0.01 event/kg/y & " & 4.1 km.w.e. & \cite{hime}& \\ 
& " &+ 95\% effective active n veto & 2.2 km.w.e. & &\\
&"&+99\% effective active n veto & 1.4 km.w.e. & & \\ \hline
ZEPLIN& $\sim$ 10$^{-10}$ pb $\Rightarrow$ 0.004 event/kg/y & 30 cm Pb, 40 g/cm$^2$ polyeth & 4.0 km.w.e. & \cite{ZEPLIN} & Boulby \\
Liquid Xe DM & " & + 95\% effective active n shield & 2.1 km.w.e & \cite{ZEPLIN} &\\ \hline
Nuclear Astrophysics & 10$^{-4}$ cts/keV/hr & LUNA $^{14}$N(p,$\gamma$) setup & 0.5 km.w.e. & \cite{LUNA} & G. Sasso \\ 
Accelerator& 2 $\times$ 10$^{-6}$ cts/keV/hr & $\times$ 50 improvement in LUNA & 2.1 km.w.e. & & \\ \hline
MAJORANA & 2.2 $\times$ 10$^{-4}$ event/keV/kg/y &granularity, PSD, segmentation & 5.0 km.w.e. & \cite{hime} &\\
$^{76}$Ge $\beta \beta$ decay & & +95\% effective active n veto & 2.9 km.w.e. & & \\
& & + 99\% effective active n veto & 2.0 km.w.e. & \cite{hime} &\\ \hline 
EXO $^{136}$Xe $\beta \beta$ decay & 10 tons $\Rightarrow$ 1.4 $\times$ 10$^{28}$ y & Ba tagging, $\phi_\mu \lsim$ 10$^{-6}$/cm$^2$ s & 1.1 km.w.e.  & \cite{EXO} & WIPP \\ \hline
CUORE  & 10$^{-3}$ event/keV/kg/y & Pb, Cu shield, anti-radon box & 3.7 km.w.e & \cite{CUORE} & G. Sasso\\
$^{130}$Te $\beta \beta$ decay& " & + 95\% effective $\mu$ anticoincidence & 1.9 km.w.e & \cite{CUORE} & \\ \hline
Radioassay & current counting levels & shielded Ge, NaI & 0.7 km.w.e. & \cite{laubenstein} & CELLAR\\
& factor-of-ten detector improvements & " & 1.8 km.w.e. & &\\ \hline
XMASSII: liquid Xe  & $^7$Be,pp $\nu$s; & self-shielding; 5cm boronic & 2.1 km.w.e & \cite{XMASS} & Kamioka \\
DM, solar $\nu$ & $\tau(\beta \beta) \sim 3 \times 10^{26}$ y & acid; ultrapure water shield & & & \\ \hline
CLEAN & 1\% cosmogenic  background  & 2900 $\nu$ events/y, $\sigma_{spall}$ $\sim$ 1 mb & 4.0 km.w.e. & \cite{mck} & \\
liquid Ne solar $\nu$  & &  & & & \\  \hline 
He TPC &  solar $\nu$s, 200 keV-2 MeV; cosmogenics & shielded TPC & 1.5 km.w.e. & \cite{bonvicini} &\\
& $\lsim$ 1\% deadtime && 2.2 km.w.e. & &\\ \hline
LENS In solar $\nu$ &  CC pp, $^7$Be, CNO $\nu$ detection & CR-induced In(p,n) $\lsim$ 10\% solar & 2.0 km.w.e. & \cite{raghavan} & \\ \hline
Water Mega-Detector & LB target, $\tau_{prot} \gsim 10^{35}$ yr , atmos. $\nu$ & fiducial volume cuts & 1.5 km.w.e & \cite{S1} & \\
& + solar $\nu$s, K$\nu$ signal & dead time $\lsim$ 10\% & 2.1 km.w.e. & \cite{SK} &  \\ \hline 
Hyper-Kamiokande & events $\gsim$ 100 MeV deposited; events & fiducial volume cuts & 1.4 km.w.e. & \cite{HK} & Tochibora\\ 
&  with timing (e.g., supernova $\nu$s) & &  & & \\ \hline
OMNIS: high Z & high statistics supernova & 8kpc: signal/noise $\sim$ 10 @ 20 s & 1.0 km.w.e. & \cite{smith} & \\
supernova-$\nu$ detector & $\nu$ light curve & 20kpc: signal/noise $\sim$ 10 @ 20 s & 1.9 km.w.e. & \cite{smith} &  \\ \hline  \hline
\end{tabular}
$^*$ Note that the minimum depths employed in the text were increased by an additional 0.6 km.w.e. in order to provide a safety factor of three.
\end{table*}

\subsection{Next-generation physics experiments that can be mounted unmodified at Stage I depths}  
A substantial fraction of the experiments listed in Table I could be mounted at Kamioka or Stage I 
depths without the use of active shields:

\noindent
$\bullet$ \textit{Nuclear accelerator for astrophysics.}   Currently a low-energy accelerator for
astrophysics, LUNA \cite{LUNA}, operates at Gran Sasso.
The LUNA $^{14}$N(p,$\gamma$) counting goal of about
10$^{-4}$/keV/hr is typical of the present state-of-the-art, which has allowed direct measurements
of red-giant and solar cross sections in the respective Gamow peaks.  Backgrounds at
this scale -- four orders of magnitude higher than next-generation double beta decay goals, for comparaison --  must come from environmental radioactivity
at Gran Sasso depths.   As the table indicates, that
rate is characteristic of cosmic ray backgrounds that prevail at $\sim$ 0.5 km.w.e.   Presumably
LUNA backgrounds are either beam associated or connected with the radiopurity of the detector
arrays now employed.  

The extent to which beam-associated and materials-associated backgrounds can be reduced is
difficult to estimate, but one would hope that substantial progress could be made with cleaner detectors and the use of passive and active shields.  But until a factor of 50 improvement comes from such
steps, cosmic-ray-induced backgrounds will not be a limiting background 
for nuclear astrophysics conducted at Kamioka depths.

\noindent
$\bullet$  \textit{Radioassay.}  The correlations of radioassay sensitivities with depth, as compiled
in Ref. \cite{laubenstein} for the Collaboration of European Low-level Underground
Laboratories, show breaks in the proportional dependence on cosmic ray backgrounds by the
depth of $\sim$ 0.7 km.w.e.  (Specifically, these results were  for high-purity Ge detectors
located in various CELLAR laboratories.) This demonstrates that radioactivity associated with materials
limits current radioassay techniques performed below this depth.   If one anticipates
a factor of ten improvement in environmental radioactivity controls in the next decade, 
this would move the
critical depth to about 1.8 km.w.e.  Thus it appears that Stage I depths would satisfy 
radioassay needs now and for a considerable period into the future.

\noindent
$\bullet$  \textit{High-energy-deposition water megadetector physics: long-baseline neutrino physics, nucleon
decay,  atmospheric neutrinos.}   Because of timing, long-baseline neutrino physics can generally
be done adequately with near-surface detectors.   However, given a site where the rock quality
is good and excavation is simple (e.g., a horizontal site with road or rail access), generally there
is little additional cost encountered in placing the megadetector at depths up to 1000m.
The advantage in such placement is the opportunity for other high-energy-deposition
physics, such as proton decay and atmospheric neutrinos.  It has been argued that depths
of 1.5 km.w.e. are adequate for such programs \cite{S1}.   A specific example is provided
by the Japanese megadetector Hyper-Kamiokande \cite{HK}, which will be sited at the Tochibora
Mine at a depth between 1.4-1.9 km.w.e.  The Hyper-Kamiokande program will focus on events
depositing $\gsim$ 100 MeV.

\noindent
$\bullet$  \textit{Water megadetector solar neutrinos.}  Super-Kamiokande has demonstrated that precise
measurements of the $^8$B spectrum down to energies $\sim$ 5 MeV are possible at
Stage I depths \cite{SK}.  The primary limitation is detector dead time, which is about 20\% for
Super-Kamiokande.   Thus if solar neutrinos are a goal, depths significantly shallower than that
provided in Stage I or at Kamioka would not be acceptable.  At the Stage II depths
discussed in the next section, the dead time would fall to  2\%.
Super-Kamiokande has also demonstrated that 
low-energy $\gamma$ rays ($\sim$ 6.3 MeV) following nuclear dexcitation could be used
as an effective tag for the K$\nu$ proton decay mode
at 2.04 km.w.e. depths \cite{SK2}.  This is an example of a  low-energy capability important
to nucleon decay in water detectors.   However, proton decay modes that produce only
low-energy events,
such as 3$\nu$ modes leading to breakup of $^{16}$O, would be observable only if
the detector is both very deep and highly instrumented.

\noindent
$\bullet$ \textit{Short-time-window physics: high-Z supernova neutrino observatories.}  Several ideas
for low-maintenance, large volume, dedicated galactic supernova neutrino detectors have been
proposed.  One proposed strategy is to exploit the large nuclear spallation cross sections
for high-Z materials such as lead, e.g., OMNIS.   Such detectors would operate very well
at Kamioka or Stage I depths because of the short time window for the burst.  A goal of
the field would be a high-statistic measurement of the $\nu$ light curve out to long times
(e.g., $\sim$ 20 s).  This would follow the protoneutron star through the period where
it radiates its lepton number, a parameter thought to control phase changes in the core 
and the conditions that might lead to late-time collapse into a black hole.   The analysis done
by Smith \cite{smith} for OMNIS in Boulby can be scaled to other depths.
One finds that cosmic ray neutron backgrounds would
be one-tenth the expected supernova signal at 20 seconds, if the detector were at 
1.0 km.w.e. and the event occurred at the galactic center (8 kpc).  The same criterion
for a supernova at 20 kpc requires a depth of 1.9 km.w.e.  The conclusion is that a
very-high statistic measurement of a galactic  supernova $\nu$ light curve would have
neglible backgrounds at Stage I depths.

\noindent
$\bullet$ \textit{Charge-current solar neutrino detection: LENS.}  The indium detector LENS \cite{raghavan}  
is one candidate for real-time charge-current measurements of the
pp, $^7$Be, and CNO-cycle neutrinos.  The principal cosmic-ray-induced background
is familiar from the chlorine experiment, secondary protons mimicking the 
solar neutrino signal by In(p,n).  This background is less severe than in the case
of Cl because the Coulomb barrier retards low-energy-proton events.  The LENS depth 
requirement has been set 
by the proposers at 2.0 km.w.e. \cite{raghavan}, based on estimates that this reduces the 
(p,n) contribution to 10\% of the solar signal.  This background would then be 
subtracted to an accuracy of 10\% by measuring the response of a prototype 
detector at lower overburdens -- a task that was also performed in the case of Cl --
leading to an overall accuracy of 1\%.    The Stage I facility meets the experiment's
requirements, and the Stage II facility discussed in the next section would reduce
the unsubtracted cosmic ray contributions to sub-1\% levels.

\noindent
$\bullet$ \textit{Neutral-current solar neutrino detection with TPC.}  The TPC 
solar neutrino experiment is focused on the spectrum between 0.2-2.0 MeV.  The
proponents have published a rather detailed background analysis \cite{bonvicini}, and have
emphasized that, in their view, site characteristics such as low excavation costs,
low radon, quality and quantity of dust control, and laboratory support are more
important to experimental success than depth, given the experiment's modest
depth requirements.  These requirements could be met by a Pioneer Tunnel
facility with ducted ventilation, good filtration systems, and rooms coated with
mineguard.   TPC is a high-pressure gas TPC that has a 30-cm outer steel
shield and an inner shield of 1.5m of hydrogen-bearing material.  The active volume
in 4000m$^3$.  Cosmic rays are used in the experiment as a calibration source.
The analysis indicates that cosmogenic backgrounds for this detector are negligible
for overburdens $\gsim$ 1.5 km.w.e.  A somewhat more stringent condition comes
from the desire to reduce dead time to a low level: the proponents recommend 
1.9 km.w.e.  The 2.2 km.w.e. from Table 1 corresponds to a dead time of 1\%.  Thus
a Stage I laboratory meets the depths requirements and would address the
proponents' construction requirements given above.

\noindent
$\bullet$ \textit{The $\beta \beta$ decay experiment EXO.}  The enriched  $^{136}$Xe $\beta \beta$
decay experiment EXO \cite{EXO} has proposed laser tagging of the daughter Ba ion to
eliminate backgrounds.  While the method has not been demonstrated, if it is
implemented, depth requirements for this experiment would be unusually modest.
One requirement, the reduction of the
muon flux to less than $10^{-6}$/cm$^2$s
to produce a tolerable false triggering rate for the laser tag of about one per hour, would
require depths in excess 1.1 km.w.e.  The proposers have generally used a more cautious 2.0 km.w.e.
as their estimate of an appropriate depth.  
The experiment is proposed for WIPP and would clearly operate well at Stage I depths.
A prototype experiment without laser tagging will be mounted later this year at WIPP.
The results will provide a quantitative basis for background assessments whether or
not Ba tagging is used.

\subsection{Depth and shielding tradeoffs for background-limited physics experiments}
Most dark matter, double beta decay, and neutral-current solar neutrino detectors have
significant background concerns.  However, in general, depth requirements have
to be viewed in the context of detector design, 
particularly shielding needs due to both environmental and cosmic ray backgrounds. 
The issues have been investigated rather carefully by experimenters proposing next-generation
detectors for Gran Sasso (3.03 km.w.e.), Boulby (2.81 km.w.e.), and 
Kamioka (2.04 km.w.e.), as well as in a recent paper by Mei and Hime \cite{hime}.

The basic issue is that a given background source, such as neutrons, may have both 
environmental and cosmogenic components, so that a shield may be a more effective
background strategy than depth.  This issue has been considered in the design and siting
of the Gran Sasso $\beta \beta$ decay experiment CUORE \cite{CUORE} and the 
Boulby dark matter experiment ZEPLIN \cite{ZEPLIN}, for example.  In each case
relatively conservative active shields (a 95\%-effective $\mu$ anti-coincidence
shield and  a 95\%-effective neutron shield, respectively) will allow these experiments
to reach aggressive goals (count rates of 10$^{-3}$/keV/kg/y and WIMP cross sections
of 10$^{-10}$ pb, respectively) at sites of intermediate depth (3.03 and 2.81 km.w.e., respectively),
while allowing some margin for error.

The DUSEL-WIPP proposal \cite{D-W} used the DRIFT dark matter detector (a gaseous CS$_2$
negative-ion time projection chamber with direction sensitivity) to illustrate the 
depth-shielding issue.  The authors found that DRIFT, if equipped with a 1-meter passive shield
and sited in a hard-rock laboratory, would be dominated by neutrons
from U/Th ($\alpha$,n) below depths of 2.6 km.w.e.  Thus placing such an experiment 
in a very deep laboratory (e.g., SNOLab) provides little benefit in the absence of 
additional shielding, despite 
the factor of 100 muon attenuation that could be gained by moving from 2.6 km.w.e. to
6.0 km.w.e.   An active shield (with at least moderate depth) is a more effective
strategy.  We believe this conclusion  holds for many similar experiments:

\noindent
$\bullet$ \textit{The dark matter and $\beta \beta$ decay experiments SuperCDMS, ZEPLIN, 
DRIFT, XENON \cite{xenon}, XMASS, and CUORE.}   Experiments like SuperCDMS 
have set very ambitious next-generation
goals, e.g., 10$^{-10}$ pb cross sections, corresponding to improvements of factors of
1600 and 30000 in the current CDMS spin-independent cross section limits for Ge and Si, respectively
\cite{superCDMS,CDMS}.
If SuperCDMS, ZEPLIN, and CUORE were conducted with passive shields similar
to those deployed in their prototypes, depths of between 4.3-4.7 km.w.e. would be needed
to suppress $\mu$-induced backgrounds to a factor of three below the design
goals of these experiments.  However, if moderately effective (95\%) active neutron (or $\mu$ anticoincidence, for CUORE)  shields are 
employed, the desired factor-of-three safety margin could be achieved at depths of
between 2.5 and 2.9 km.w.e.  At Stage I depths (2.12 km.w.e.) muon-associated backgrounds would be
expected at about the dark-matter design sensitivity of $10^{-10}$ pb.
These observations are consistent with the proposed siting
of ZEPLIN at Boulby and CUORE at Gran Sasso.

Given a sufficient investment in shielding, these experiments would have discovery potential
at their design goals -- that is, signals above background -- if
mounted in a Stage I laboratory.  If active shields that are 99\% effective are employed, the
experiments could be mounted at depths of 1.5-1.8 km.w.e. with backgrounds no more than
one-third their design goals.

XMASS \cite{XMASS} is an example of a dark matter (and solar neutrino) experiment designed for
Stage I depths (Kamioka).   It includes an
ultrapure water shield, a neutron shield consisting of 5 cm of boronic acid, and a
detector that will allow fiducial volume cuts.  It is currently in a prototype stage in which
the effectiveness of the shielding strategy is being evaluated. 

\noindent
$\bullet$ \textit{MAJORANA.}  The $^{76}$Ge $\beta \beta$ experiment MAJORANA \cite{majorana} has
the design goal of 2.2 $\times$ 10$^{-4}$ events/keV/kg/y.   The proposed design
exploits detector granularity, pulse-shape discrimination, and detector segmentation
to reduce background, and assumes an active muon veto effective at 90\%.   While
in principle the addition of a 99\%-effective active neutron shield would allow the
experiment to be mounted in the Stage I laboratory, at this depth $\mu$-induced 
background would be at the level of the experiment's counting goal.  Unlike the dark
matter experiments described above, no error margin would be available. 

\noindent
$\bullet$ \textit{Neutral current solar neutrino detectors.}
Large-volume neutral current detectors designed to measure pp and $^7$Be solar neutrinos
may need to be sited quite deep.
CLEAN \cite{mck}, a liquid Ne detector in the R\&D phase, is the example included in Table 1,
though the liquid He experiment HERON \cite{heron} would be an equally good choice.
As one of the goals of CLEAN is a 1\% measurement of the pp $\nu$ flux, the minimum
depth given in Table 1 was determined by demanding that the cosmogenic background
not exceed this level.  This calculation depends on poorly known spallation cross sections,
several of which were discussed in Ref. \cite{mck}.  Troublesome activities are those
that produce events in the observation window and have long lifetimes, so that they
cannot be correlated with a muon and vetoed, e.g., $^7$Be.
The assumed spallation cross section for this isotope  of $\sim$ 1mb \cite{mck} 
leads to a depth requirement of 4.6 km.w.e. -- though the adopted cross section could easily be an
order of magnitude too large.  

Neutral-current neutrino detectors generally require significant depth: 
similar considerations led to placing SNO at 6.0 km.w.e.  In water detectors important 
spallation products include $^{12}$N, $^{12}$B, $^8$B, and $^8$Li, 
with the $\beta$s delayed from 10 to 800 ms.  Had SNO had been mounted at Gran Sasso depths, 
the vetoing of such activities would have produced a detector
dead time of 40\% (assuming no adjustment of other cuts) \cite{nrc}. 

In this context, we note that a proposed followup to SNO -- the SNO Liquid Scintillator
Project -- has been proposed to measure the pep solar neutrino line \cite{snoplus}.
The very great depth of SNO is crucial to this experiment, as otherwise
cosmogenic production of $^{11}$C in carbon-based scintillator would obscure
the signal.

\subsection{Summary of depth requirements for next-generation experiments}
{\it We conclude, from the discussion above, that SNOLab and a Stage I laboratory would meet
the needs of the experiments likely to be mounted in the next decade.}  The experiments would seem
to break into four classes:

\noindent
$\bullet$ Large classes of next-generation experiments can be conducted successfully and
economically in a Stage I laboratory, without changes in design.  They include accelerator
experiments for nuclear astrophysics; standard radioassay, including virtually all counting
important to national security; high-energy-deposition megadetector
experiments such as long-baseline $\nu$ oscillations, nucleon decay, and atmospheric
neutrinos; megadetector studies of $^8$B solar $\nu$s;  large observatories for measuring
the supernova $\nu$ light curve out to long times; certain charge-current and neutral-current
solar neutrino detectors such as LENS and TPC; and certain $\beta \beta$ decay experiments
such as EXO.

\noindent
$\bullet$ Next-generation dark matter experiments such as SuperCDMS and ZEPLIN III and 
$\beta \beta$ decay experiments such as CUORE, if mounted at Stage I depths with 
active vetos that are 95\% effective, would expect to see cosmogenic backgrounds at about
their design sensitivities (10$^{-10}$ pb or 10$^{-3}$ events/keV/kg/y, respectively).  
Such vetos would likely be important, regardless of depth, to suppress radioactivity backgrounds.
Vetos that are 99\% effective would provide enough additional suppression to allow discovery
at the design sensitivities.   Thus a great deal could be accomplished at Stage I depths,
particularly considering the large gap between the dark-matter 10$^{-10}$ pb goal and
current limits.

\noindent
$\bullet$  At Stage I depths MAJORANA, if equipped with a 95\% effective active neutron veto, would
expect to see cosmic ray-associated neutrons at about the rate 6 $\times$ 10$^{-4}$ events/keV/kg/y,
about three times the design goal \cite{hime}.  The experiment in principle 
could reach its design goal with
a 99\% effective active neutron veto in the Stage I laboratory, but with little margin. 
 
\noindent
$\bullet$ Experiments such as CLEAN, a neutral current detector for low-energy solar neutrinos,
require depths of about 4.5 km.w.e. because of long-lived cosmogenic activities.  
SNOLab is the one North American laboratory providing such depth..

We conclude that major portions of the North American underground science program could be
conducted very successfully at Stage I depths.  In many cases experiments would be more
successful if
mounted at the Stage I depths because of the Pioneer tunnel's superior access, lower construction
costs, and ease of excavation.  Stage I would clearly be adequate for activities 
like low-level counting for national security, an effort where a U.S. site might be mandatory. 

A few experiments, to reach their design goals, would require deep space, such as that available
at SNOLab.  It is likely that the U.S. will elect to pursue a double beta decay experiment like
Majorana, a dark matter experiment like SuperCDMS, and a low-energy neutral current solar
neutrino detector, like CLEAN or HERON, some time in the next decade.  SNOLab has adequate 
space for these experiments.  In our view, an experiment like SuperCDMS could also be successful
at Stage I depths, with suitable design of active shields.

We conclude that a cooperative US-Canadian program for North American underground science
in which experiments are sited in either SNOLab or Stage I, according to their needs, would be
a cost effective strategy for mounting next-generation experiments.  

\subsection{Geoscience and Geomicrobiology}
Geoscience and geomicrobiology have played a prominent role in DUSEL discussions.
While a discussion of this science is beyond the scope of this paper, here we list 
physical attributes of the site that would be relevant to a crystalline rock geosciences
program.   In Stage I the primary feature of interest to geoscience is the
five kilometers of unlined ventilation/utility tunnel that would be tracked and lighted.  The 
surface above the tunnel is accessible due to Stevens Pass Ski Area jeep trails, which opens up
possibilities for geophysical measurements from the surface to sensors at depth and for verification
core drilling.

The area's thermal behavior has been studied because of geothermal activity near the Scenic portal. 
Borehole measurements have mapped out local regions of elevated
gradients ranging to 68$^\mathrm{o}$C/km \cite{schuster}; these studies were used in
thermal modeling of the batholith.  Much of the 
batholith has been recently altered by glaciers which cut into the mesophile zone.  Thus
geomicrobiology issues would include the influence of geothermal activity, recent rapid cooling,
and geochemical weathering on microbial migration. 

Serpentenite bodies have been mapped within the batholith.  Depending
on the results of more detailed geologic surveys of the Pioneer tunnel, there may be opportunities
to evaluate ophiolite weathering as a local source of H$_2$ for sustaining microbial activity.

The surface access would allow studies of groundwater flux between shallow and deep flow
systems, global climate change implied by groundwater geochemistry and isotopics, the
hydrology of fracture flow systems, and potentially soil microbiology issues associated with
glacial retreat in the batholith.  These studies would help determine the relatedness of
surface and near-surface microbial communities and deep communities.

Were Stage II to be undertaken, there would be opportunities to study transient processes
during tunneling, to evaluate numerical predictions for fractured media, and to image fractured
rock systems.

Geology and tectonics issues include the emplacement mechanisms for plutons, determining
whether large plutons have floors, and direct analysis of paleomagnetic uncertainties and cooling
effects.

Finally, the 5 km of unlined tunnel provides access to granodiorite and schist that could serve
as host to large-scale, long-term geoscience experiments, such as the creation of a synthetic
ore deposit \cite{S1}.  Such an experiment would require limited additional excavation south
of the Pioneer tunnel, away from railway activities.  Combined with geophysical monitoring,
it would provide a remarkable opportunity for solving long-standing problems regarding
scaling of key ore-generation processes.  Similarly, manipulation of subsurface fluid flux
and stress conditions (thermal or mechanical) will permit testing of recently developed 
theories for scaling of rock mass behavior by direct observation and measurement.  Some
of these experiments can be done as part of Stage I excavations of hallways and experimental
rooms, but others will require specialized excavations to support scientific objectives.

\section{Stages II and III}
The analysis of the previous section suggests that the combination of SNOLab and Stage I
would be an effective strategy for providing underground space for 
the experiments North American scientists are currently considering.
As these sites have complementary strengths, cooperation between
the laboratories would be natural.   Both sites have low operations costs, so that underground
science funding could be focused on the experiments that these sites (and Soudan and WIPP) 
will house.  However, underground science is a growing field, important not only to physics
but also to earth science, engineering, microbiology, and materials science.  This growth
could soon require more deep space.
One of the important attributes of the Pioneer tunnel site is its potential for providing that
space through cost-effective upgrades.

\begin{figure*}
\includegraphics[width=16cm]{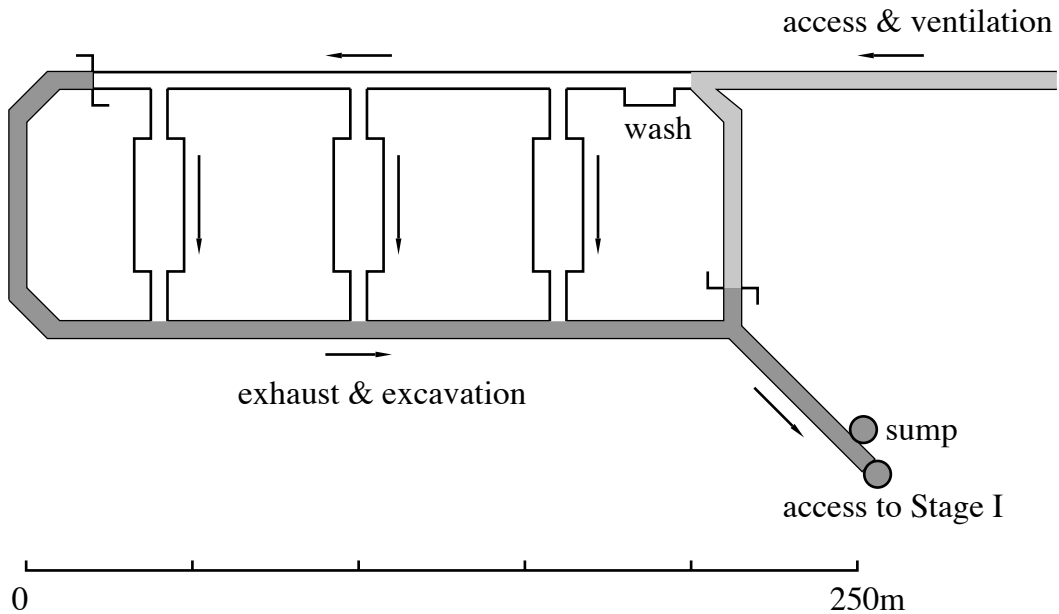}
\caption{A schematic of the envisioned Stage II laboratory.  The shading indicating 
ventilation flow is as described in Fig. 9.  Stages I and II would
be connected by a 560m shaft that would allow Stage II to make use of the exhaust system,
power, drainage, and haulage systems of Stage I.}
\end{figure*}

While Stage II could be built in parallel with Stage I, it would be attractive to delay this
stage (if the need for space beyond SNOLab and Stage I is not urgent) so that its
design can be influenced by the experience gained from operating Stage I.  This
experience would include a better understanding, on the part of the scientific community,
of the needs of next-to-next-generation experiments that might be undertaken in
Stage II.  Stage II would be able to house experiments needing significant depth.

Laboratory staging offers many advantages.  It allows facility development to proceed
at pace with experimental development.   It saves money by avoiding
large up-front facility investments and associated operations
costs.  In the case of the Pioneer tunnel plan, the staging is designed to maintain a
complementary relationship with SNOLab, giving North America new capabilities but
not competing with SNOLab for very great depth.   We envision a very deep Stage III
only if SNOLab closes.

The Stage II concept sketched here shows how one can
build a deeper, horizontal-access laboratory by utilizing much of the
investment made in Stage I, e.g., the Stage I tunnel enlargement, the drainage system, 
the electrical system, and the tracked haulage system.   The new excavation required
is still quite modest on DUSEL standards, 4.8 km of entrance tunnel and a 560m 
5m-diameter shaft linking Stages I and II (which could be built quickly and economically 
by the raise-bore method).  All excavation is confined to the immediate vicinity of the
Pioneer tunnel -- rock with a well-understood geology and a track record for favorable
construction.  Despite its depth, all of this rock would be easily cored because we will have
the Pioneer tunnel and laboratory as a convenient, deep location from which to drill.
That is, construction risks and exploration costs would both be low.

Stage II would be an exceptionally clean, horizontal access laboratory with a depth of
3.62 km.w.e. (and a peak overburden of 4.65 km.w.e.), a depth that is intermediate between
Gran Sasso (3.03 km.w.e.) and Frejus (4.15 km.w.e.), and similar to the proposed
DUSEL-Henderson central campus (6750 ft elevation, 3.81 km.w.e.).   Assuming
modern methods could match the 1920's advance rate of 11m/day (a good rate
even by today's standards), excavation of the new
access tunnel could be completed in about one year.  Thus the Stage II laboratory 
could be established quickly.

Stage II would begin with the excavation of a -10\% downgrade tunnel from the Scenic portal
to reach the region of maximum overburden near Cowboy Mt., shown in Fig. 7.  This grade
is compatible with rubber-tire access, and has been used on European public roadway tunnels
with good safety records.  A detailed discussion of tunnel construction and finishing is
available in the DUSEL-Cascades proposal \cite{DC}.  A tunnel 
of this length is a candidate for either tunnel-boring-machine or drill-and-blast excavation.
Haulage would be done by rail.
Note that the effective gain in depth relative to Stage I is 11.6\%, due to the grade of the
Pioneer tunnel.

The purpose of the new tunnel is to reach a location 560m below Stage
I, but aligned with Stage I in such a way that Stage I facilities can support Stage II.
The new tunnel would serve as the access tunnel and ventilation intake for Stage II.   Clean air
could be brought into the new tunnel at the Tye Incline and directed to the Stage II laboratory.
Experimental areas would be behind a clean barrier/car wash, as in Stage I.
As shown in
Fig. 10, a concrete-lined raised-bore shaft would then be constructed as a
 ``dirty-side-to-dirty-side" connection between Stages I and II.
The sump for the Stage II level would be located near the 
base, with a pump column in the shaft linking the sump to the Pioneer tunnel drain system.
The new tunnel and Stage II would drain by gravity to this sump and pump column.
The independent distribution lines to Stage I would be extended down the shaft,
to provide power to Stage II.  The shaft would carry exhaust from Stage II up to Stage I,
where it would join the ``dirty-side" flow out the Stage I entrance tunnel -- which now becomes
the exhaust path for both stages.

The shaft would be equipped with a winze and would serve as the secondary escape for
Stage II, as a route for quickly transporting personnel and small equipment between Stages I and II,
and as a mining hoist for Stage II.  Visitors could park at Stage I, take the elevator to
Stage II, and enter the laboratory through the Stage II
clean barrier/car wash.  Once the new tunnel is completed and Stage II
established, all future Stage II excavation
would be conducted via this shaft, in an effort to keep the new tunnel
and the experimental areas on Stage II as clean as possible.  Stage II excavation would
utilize the Stage I tracked haulage system.   
As the shaft is relatively short,
a modest hoist could provide Stage II with significant mining capacity.

We envision Stage III -- a full DUSEL -- occurring only in the far future, and only if some
event like the closure of SNOLab leads to a shortage of very deep space.
A third deeper level, aligned vertically with Stages I and II, could be established
by spiraling downward from Stage II, followed by extension
of the raised-bore shaft to that level.   Such ramps have been used in other hard-rock 
sites.  An additional 4.8 km of tunneling would be needed to reach 5.0 km.w.e., if the -10\% gradient
were continued.  

Stages II and III would provide increased access to pristine rock at higher temperatures and stress,
enhancing research opportunities in geomicrobiology, stability of mined openings, and
geochemical flux of dissolved constituents through fractures in the granodiorite as a
function of depth.

\section{Conclusion}
The analysis presented here was stimulated by BNSF's recent expression of interest in the scientific use
of the Pioneer tunnel.   This tunnel provides a rare opportunity to establish a dedicated Stage I
horizontal-access laboratory by exploiting an existing opening in high quality rock.   Development
risks are low, construction costs are modest, and operations costs -- which often dominate
lifetime project costs -- would be exceptionally low.  The site
could be developed to provide separate entrance, ventilation, and escape tunnels, features
one would generally expect only in a laboratory excavated for science.   The site's advantages
include private ownership, existing permits for drainage and ventilation, 
excellent railroad and highway access, proximity to
a major metropolitan area and international airport, stable and inexpensive power,
a rock temperature at the laboratory site of 21$^\mathrm{o}$ C,
and good potential for staged development.  The site is also ideal in complementing 
international efforts toward a neutrino factory, with a ``doubly magic" baseline (CERN, KEK)
and a separation from FermiLab that is close to optimal for neutrino-factory CP-violation
studies.

The site presents an opportunity to quickly establish a Stage I horizontal-access 
laboratory that would be similar to Kamioka in depth.
As the only North American drive-in laboratory, and the
only laboratory with dedicated portal-to-laboratory clean access,
it would play a unique role.  The laboratory would be an outstanding complement
to Canada's deep laboratory SNOLab, which will be completed
by the end of 2007, as well as to the two existing US vertical facilities.
SNOLab and Stage I, by
working together, could meet the needs of the experiments now 
under consideration.   Experimenters could select a site,
depending on the requirements most important to the experiment:
depth, ease of access, cleanliness, excavation capability, etc.  
A  partnership between SNOLab and the U.S. -- ideally one that includes WIPP and
Soudan, in addition to Stage I --  would appear to provide a low
cost solution to our facilities needs,
conserving available funding for the many
new experiments now waiting in the R\&D phase.

Stage I offers attractive opportunities for geoscience and geomicrobiology research, providing
otherwise unavailable access to extensive subsurface rock exposures and deep pore
and fracture fluids.  Such access also will enable long-term geochemical and geomechanical
experiments involving induced thermal, flux, and mechanical stresses on rocks
under near {\it in situ} conditions.

Stage I also addresses important efficiency, safety, and security issues affecting one
of the nation's key transportation corridors, and the most critical choke-point for
the nation's container traffic.  These issues are important to possible partners.

We have argued that the combination of SNOLab and Stage I would meet the requirement
of the next-generation experiments planned for the next decade.  However, one of the
attractive features of a long-term joint Canadian/U.S. underground science partnership is
the opportunity to respond effectively to future needs, with the U.S. upgrading its facility 
in stages.  Staging minimizes risk because new facilities can be designed and built
in response to emerging scientific needs, taking advantage of the knowledge gained in
constructing and operating an early stage or stages.
Staging is cost effective because it delays large facility investments until
they are needed, and avoids unnecessary operations costs. 
We envision Stage II (3.62 km.w.e.) being undertaken near the end of the next decade, building
on the infrastructure improvements made for Stage I.
It would give the U.S. an enlarged, deeper, exceptionally clean laboratory with
dedicated horizontal-access.  It would be designed to complement SNOLab.
Stage I/Stage II and SNOLab (6.0 km.w.e.) would be an effective answer to the
joint European laboratory, the partnership between Gran Sasso (3.03 km.w.e.)
and a much enlarged Frejus (4.15 km.w.e.), and other world-leading
facilities.  This is a cost effective, pragmatic strategy
for strengthening North American underground science while maximizing the 
amount of science we can do in the next two decades.

\acknowledgments
We  thank Tom Gaisser and Todor Stanev for their advice on
muon production and attenuation; Marvin Marshak, Hank Sobel, and Raju Raghavan
for comments; and David Gilliam, Lee Grodzins,
Dick Kouzes, and Jeff Nico for
discussions on the cargo container issue. 
This work was supported by the U.S. Department of Energy, Office of
Nuclear Physics and
grants DE-FG02-00ER-41132.

\end{document}